\documentclass[10pt]{iopart}
\usepackage{hyperref}
\usepackage{amssymb}
\usepackage{amsthm}
\usepackage{enumerate}
\usepackage{bm}
\usepackage{graphicx}
\usepackage{float}
\usepackage[small, nohug]{diagrams}
\diagramstyle[labelstyle=\scriptstyle]

\usepackage{xcolor}
\usepackage{amssymb}
\usepackage{bm}

\usepackage{graphicx}
\usepackage{amsmath,amsthm,amsfonts,amscd}
\usepackage{amssymb}
\usepackage{url}
\usepackage{mathtools}
\usepackage[all]{xy}
\usepackage{slashed}
\usepackage{multirow}
\usepackage{hyperref}

\newcommand{\alg}{{\mathcal A}}
\newcommand{\sa}{{\mathcal S \mathcal A}}

\newcommand{\ha}{\widehat{\mathcal A}}
\newcommand{\hbA}{\bm{{\widehat \Aa}}}
\newcommand{\tbA}{\bm{{\widetilde \Aa}}}
\newcommand{\tbT}{\bm{{\widetilde \Tt}}}

\newcommand{\wt}[1]{\widetilde{#1}}
\newcommand{\wh}[1]{\widehat{#1}}

\newtheorem{theorem}{Theorem}

\newtheorem{proposition}[theorem]{Proposition}
\newtheorem{corollary}[theorem]{Corollary}
\newtheorem{definition}[theorem]{Definition}
\newtheorem{remark}[theorem]{Remark}
\newtheorem{example}[theorem]{Example}

\newcommand{\CM}{{\mathbb C}}
\newcommand{\NM}{{\mathbb N}}
\newcommand{\RM}{{\mathbb R}}
\newcommand{\SM}{{\mathbb S}}
\newcommand{\TM}{{\mathbb T}}
\newcommand{\ZM}{{\mathbb Z}}

\newcommand{\KM}{{\mathbb K}}

\newcommand{\UM}{{\mathbb U}}

\newcommand{\Aa}{{\mathcal A}}

\newcommand{\Ss}{{\mathcal S}}

\newcommand{\Tt}{{\mathcal T}}

\newcommand{\Nn}{{\mathcal N}}
\newcommand{\Mm}{{\mathcal M}}

\newcommand{\Jj}{{\mathcal J}}

\newcommand{\Hh}{{\mathcal H}}

\begin{document}

\title[Bulk-Boundary Correspondence for Topological Insulators]{Bulk-Boundary Correspondence for Topological Insulators with Quantized Magneto-Electric Effect}

\author{Bryan Leung}

\address{Magnetic, New York, NY, U.S.A., email: bryleung@berkeley.edu} 

\author{Emil Prodan}
\address{Department of Physics and Mathematical Sciences,
Yeshiva University, U.S.A., email: prodan@yu.edu}

\date{}

\begin{abstract}
We study bulk-boundary correspondences and related surface phenomena stabilized by the second Chern number in three-dimensional insulators driven in adiabatic cycles. Magnetic fields and disorder effects are incorporated in our analysis using operator algebraic methods. We use the connecting maps between the $K$-theories of bulk and boundary algebras as engines for the bulk-boundary correspondences. We discovered that both the exponential and the index connecting maps are relevant for the context considered here as they lead to distinct experimentally observable surface phenomena, such as pumping and transfer of quantum surface Hall states or proximity induced Hall effect. The surface Hall physics of time-reversal symmetric topological insulators is also investigated using the new tools, which can model irrational magnetic fluxes and arbitrary large surface disorder. 

\end{abstract}

\maketitle

{\scriptsize \tableofcontents}

\setcounter{tocdepth}{2}
\numberwithin{equation}{section}
\numberwithin{theorem}{section}

\section{Introduction}

The total variation $\Delta \vec P$ of the electric polarization vector during an adiabatic cycle of a gapped three-dimensional (3D) lattice Hamiltonian $\{H(t)\}_{t \in \SM^1}$ can be computed as \cite{ST2013}
\begin{equation}\label{Eq:TP}
\Delta P_i = \int_{\SM^1} {\rm d} t \, \Tt \big (\bm P_G \big [\partial_0 \bm P_G,  [X_i, \bm P_G] \big]\big ), \ \ i=1,2,3,
\end{equation}
where $\Tt$ is the trace per volume, $\bm P_G := \{P_G(t)\}_{t \in \SM^1}$ is the collection of spectral projections of the family $H(t)$ onto the spectrum below the gap $G$, $\partial_0$ is the derivation with respect to time and $\bm X=(X_1,X_2,X_3)$ is the position operator. The above expression is particularly interesting because the right hand side coincides with the weak topological Chern invariants that can be associated to $\bm P_G$ \cite{PS}. Going one step further, the magneto-electric response tensor is defined as $\alpha_{i,j}=\partial P_i/\partial B_j$ where $\bm B$ is an externally applied magnetic field. The net changes of the diagonal components $\alpha_{i,i}$ under an adiabatic cycle have been computed in \cite{LP, PS} and can be expressed as
\begin{equation}\label{Eq:TME}
\Delta \alpha_{i,i} =  4 \pi \imath \int_{\SM^1} {\rm d}t  \, \Tt \Big (\bm P_G  \big [ \partial_0 \bm P_G ,  [ X_i, \bm P_G ] \big] \big[ [X_j, \bm P_G ] , [X_k, \bm P_G] \big]\Big), \ \ i=1,2,3,
\end{equation}
where $j,k \in \{1,2,3\}$ are chosen such that $ i \neq j \neq k$. The above relation shows that the total variations of the diagonal elements $\Delta \alpha_{i,i}$ are all equal and this common value will be denoted as $\Delta \alpha_{\rm ME}$. One should note that, previously $\Delta \alpha_{\rm ME}$ was defined as the average of $\alpha_{i,i}$'s, but \cite{PS} showed that such average is not necessary, a point that might prove to be essential for the experimental measurements. On the right hand side of \eqref{Eq:TME}, one can recognize the unique strong Chern invariant associated to $\bm P_G$. As opposed to its weak counterparts, this strong invariant was proven to be quantized and robust against strong disorder \cite{BES,PLB}. The present study investigates the surface physics induced by a non-trivial quantized value of $\Delta \alpha_{\rm ME} $. 

\vspace{0.2cm}

For a 3D time-reversal symmetric (TRS) topological insulator, $\Delta \alpha_{\rm ME} $ is an odd number whenever the adiabatic cycle is mapped onto itself by the time-reversal operation and when the cycle connects the topological insulator with a trivial one \cite{LP}. In particular, it cannot be zero. The topological magneto-electric effect emerged as one of the fundamental characteristics of 3D TRS topological insulators \cite{QHZ2008, EMV2009, ETMV2010, MSCV2010, OTVS2017} and several experimental manifestations of it were proposed \cite{QLZZ2009, TM2010, MQDZ2010, NN2011,MFN2015}. Experimental efforts towards a direct observation of the quantized magneto-electric have been undertaken \cite{XJS2018, MKT2017} but the goal remains elusive. Topological magneto-electric response was also predicted to have measurable quantum magneto-optical effects, specifically via Faraday and Kerr rotations \cite{TM2010, MQDZ2010}, and these effects were later measured in topological insulator thin films \cite{Okada2016, Wu2016, DSP2017}.  In addition, quantized magneto-electric effects exist in many other classes of 3D topological phases, including inversion-symmetry protected insulators \cite{LL} and glide-symmetric crystalline insulators \cite{VdJL}. The surface physics of 3D TRS topological insulators and of the others just mentioned is quite rich, precisely due to the topological bulk magneto-electric responses. For example, when time-reversal symmetry is broken at the surfaces, quantized Hall effect can occur even in the absence of externally applied magnetic fields. The quantized anomalous Hall effect was predicted to exist when the surface of a topological insulator is decorated with magnetic atoms \cite{YZZZ2010} and the prediction was experimentally confirmed \cite{Chang2013, CYTT2014}. The Landau quantization of topological surface states under an externally applied magnetic field was observed in Bi$_2$Se$_3$ \cite{Cheng2010, Hanaguri2010}. Later, a ladder of odd integer quantum Hall plateaus have been observed in topological thin films under magnetic fields \cite{Zhang2014, Yoshimi2015}, providing evidences of half-integer Hall quantization for the individual surfaces of the film.

\vspace{0.2cm}

In this work, we identify the topological quantized surface effects for adiabatically driven 3D insulators using the connecting maps in complex $K$-theory. These maps allow for a rigorous and comprehensive connection between the topology of bulk and boundary observables. For example, the explicit computation of these maps enable us to identify and describe several interesting experimental settings where unique physical effects can be observed under cyclic adiabatic deformations. Some of these effects are completely overlooked by the physics literature. One of them is the pumping of states between the bulk valence and conduction bands. These pumped states carry quantized Hall effect. For example, a conduction band, which initially was trivial, can display quantized Hall effect in our conditions of pumping. Another prediction is that, if the surfaces of two adiabatically driven insulators are brought in close proximity, Hall states can be transferred form one surface to the other in these conditions of pumping.

\vspace{0.2cm}

We end our introductory remarks with a few words about the $K$-theoretic formalism, which was initiated in the pioneering works \cite{SKR, KRS} where the bulk-boundary problem for the integer quantum Hall effect was solved. Following this model, the $K$-theoretic formalism has been recently extended to cover the entire classification table of topological insulators \cite{PS, BKR}. It has been also applied to many quasi-periodic and aperiodic systems \cite{KellendonkAHP2019,ProdanJGP2019}. Another important development is an application of the formalism to Floquet topological insulators \cite{Sadel2017}. There are similarities but also differences in the computations performed in \cite{Sadel2017} and our study. For example, the boundary-maps involving the suspension algebra are initiated from $K_1$-group in \cite{Sadel2017}, while in our case from $K_0$-group. The physical interpretation and predictions are also of completely different nature in these two studies.

\section{Formalism}\label{Sec:Formalism}

In this section, we introduce generic bulk, half-space and boundary Hamiltonians and their associated $C^{\ast}$-algebras. They model physical systems with arbitrary magnetic fields and disorder. Time-reversal symmetric topological insulators are treated separately. We also introduce the bulk, half-space and boundary $C^\ast$-algebras for adiabatically driven systems. Furthermore, Chern numbers are defined for the algebras of bulk and boundary observables in the dynamical setting.

\subsection{Bulk physical observables}

We study covariant families of lattice Hamiltonians $H_\omega$ indexed by a configuration space $\Omega$. The space $\Omega$ is assumed to be contractible and is endowed with
\begin{enumerate}[{\rm i)}]
\item a compact Hausdorff topology;
\item an action of lattice translations, {\it i.e.} a group homomorphism $\tau: \ZM^3 \rightarrow {\rm Homeo}(\Omega)$;
\item a probability measure ${\rm d}\mathbb P$, which is invariant and ergodic with respect to $\tau$.
\end{enumerate} 

\begin{remark}{\rm The probability measure $\mathbb P$ on $\Omega$ represents the Gibbs measure, which comes from the atomic degrees of freedom. For pure homogeneous thermodynamics phases, the Gibbs measure is indeed $\tau$-invariant and ergodic. Among other things, this ensures that the intensive thermodynamic coefficients, obtained as traces per volume, do not fluctuate from one configuration to another. Also, the spectra ${\rm Spec}(H_{\omega})$ are $\mathbb P$-almost surely independent of $\omega$. First-principles computations of the Gibbs measure for the silicon crystal at various temperatures can be found in \cite{KuhneProdan2017}.
}$\Diamond$
\end{remark}

\begin{example}{\rm  For disordered Hamiltonians with white noise, $\Omega$ consists of the Hilbert cube $\big[-\tfrac{1}{2},\tfrac{1}{2}\big]^{\times \ZM^3}$ equipped with product topology and probability measure 
\begin{equation}
{\rm d}\mathbb P(\omega) = \prod_{\bm x\in \ZM^3} {\rm d}\omega_{\bm x}, \quad \omega = \{\omega_{\bm x}\}_{\bm x \in \ZM^3} \in \Omega.
\end{equation} 
The lattice translations act as $\tau_{\bm y}\{\omega_{\bm x}\}_{\bm x \in \ZM^3} = \{\omega'_{\bm x}\}_{\bm x \in \ZM^3}$, with $\omega'_{\bm x}=\omega_{\bm x-\bm y}$.
}$\Diamond$
\end{example} 

\vspace{0.2cm}

The presence of a magnetic field is captured by the flux matrix $\Phi = \{\phi_{ij}\}_{i,j=\overline{1,3}}$ and the covariant property reads
\begin{equation}\label{Eq:CovariantProp}
V_{\bm x} H_\omega V_{\bm x}^\ast = H_{\tau_{\bm x} \omega}, \quad \bm x \in \ZM^3,
\end{equation}
where the $V$ operators represent the magnetic lattice translations
\begin{equation}
V_{\bm x} = e^{-\imath \langle \bm x, \Phi \bm X \rangle } S_{\bm x}, \quad V_{\bm x}V_{\bm y} = e^{-\imath \langle \bm x,\Phi \bm y \rangle } V_{\bm y} V_{\bm x}, \quad V_{\bm x}^\ast = V_{\bm x}^{-1},
\end{equation}
with $S_{\bm x}$ the shift operators $S_{\bm x}|\bm y\rangle = |\bm y + \bm x \rangle$ and $\langle \cdot,\cdot \rangle$ the Euclidean scalar product. Taking into account for $N$ internal degrees of freedom, the Hilbert space of the physical models is $\Hh =\CM^{N} \otimes \ell^2(\ZM^3)$ and the generic covariant Hamiltonians take the form
\begin{equation}\label{Eq:GenHam}
H_{\omega} = \sum_{\bm x \in \ZM^3}\sum_{\bm y \in \ZM^3} h_{\bm x}(\tau_{\bm y} \omega) \otimes |\bm y\rangle \langle \bm y |U_{\bm x}, \quad h_{\bm x}^\dagger (\omega) = h_{- \bm x} (\omega).
\end{equation}
Here, $h_{\bm x} (\omega)$ belongs to $C_N(\Omega)$, the algebra of continuous functions over $\Omega$ with values in the algebra $M_N(\CM)$ of $N \times N$ matrices with complex entries, and $U_{\bm x}$'s are the dual magnetic translations
\begin{equation}
U_{\bm x} = e^{\imath \langle \bm x, \Phi \bm X \rangle } S_{\bm x}, \quad U_{\bm x}U_{\bm y} = e^{\imath \langle \bm x,\Phi \bm y\rangle } U_{\bm y} U_{\bm x}, \quad U_{\bm x}^\ast = U_{\bm x}^{-1}.
\end{equation}
The sum over $\bm x$ in \eqref{Eq:GenHam} runs over a finite subset $\mathcal{R}$ of $\ZM^3$ to ensure a finite range hopping of the Hamiltonians. The set of $H_{\omega}$ satisfying $\eqref{Eq:CovariantProp}$ defines the covariant family of Hamiltonians $H = \{ H_{\omega} \}_{\omega \in \Omega} $.

\begin{example}\label{Ex:FirstNN}{\rm The magnetic lattice Hamiltonians can be modeled via Peierls substitution \cite{Peierls1933}. For example, the first-nearest neighbor hopping Hamiltonian $H = \sum_{<\bm x,\bm y>} |\bm x\rangle \langle \bm y|$ becomes $H = \sum_{<\bm x,\bm y>} e^{\imath \langle \bm x, \Phi \bm y \rangle} |\bm x\rangle \langle \bm y|$ in the presence of a magnetic field. It is straightforward to see that $H$ can be recast in the form $H = \sum_{|\bm x|=1} U_{\bm x}$.
}$\Diamond$
\end{example}

\vspace{0.2cm}

The generic covariant Hamiltonians \eqref{Eq:GenHam} can be generated from the universal $C^\ast$-algebra
\begin{equation}
\Aa_3= C^\ast \big (C_N(\Omega),u_1,u_2,u_3 \big ),
\end{equation} 
defined by the commutation relations
\begin{equation}\label{Eq-BulkCommRel}
u_i^\ast u_i = u_i u_i^\ast = 1, \ u_i u_j = e^{\imath \phi_{ij}}u_j u_i, \ fu_j=u_j(f\circ \tau_j), \quad f \in C_N(\Omega),
\end{equation}
where $\tau_j$'s are the actions of the generators of $\ZM^3$ on $\Omega$. A generic element from $\Aa_3$ takes the form
\begin{equation}
a = \sum_{\bm x \in \ZM^3} a_{\bm x} \, u_{\bm x}, \quad a_{\bm x} \in C_N(\Omega), \quad u_{\bm x} = u_1^{x_1} u_2^{x_2} u_3^{x_3},
\end{equation}
where the coefficients $a_{\bm x}$ manifest a certain decay with $\bm x$ \cite{PS}. The algebra admits a family of canonical traces supplied by the point evaluations
\begin{equation}
\Tt_{\omega}(a) = {\rm tr} \big ( a_{\bm 0}(\omega)\big), \quad a \in \Aa_3,
\end{equation}
whose GNS-representations supply a family of canonical representations
\begin{equation}\label{Eq-BulkRep}
C_N(\Omega) \ni f \rightarrow \pi_\omega(f) = \sum_{\bm x \in \ZM^3} f(\tau_{\bm x}\omega) \otimes |\bm x \rangle \langle \bm x |, \quad \pi_\omega (u_{\bm x}) = U_{\bm x}, 
\end{equation}
such that $h=\sum_{\bm x}h_{\bm x} \, u_{\bm x}\in \Aa_3$ generates \eqref{Eq:GenHam} as its operator representation. 

\begin{example}{\rm Let $h=\sum_{|\bm x|=1} u_{\bm x}$. Then $\pi_\omega(h)$ is precisely the Hamiltonian introduced in Example~\ref{Ex:FirstNN}.
}$\Diamond$
\end{example}

\vspace{0.2cm}

A non-commutative differential calculus can be put in place for the algebra $\Aa_3$. Indeed, the commutation relations \eqref{Eq-BulkCommRel} are invariant to $U(1)$-twists of the generators $u_i$ and, as such, the algebra accepts a continuous action of the 3-torus $\rho:\TM^3 \rightarrow {\rm  Aut}(\Aa_3)$ given by
\begin{equation}
\rho_{\bm \lambda}(u_j) = \lambda_j u_j, \quad {\bm \lambda} =(\lambda_1, \lambda_2, \lambda_3)\in \TM^3,
\end{equation}
whose generators provide three derivations $\partial_j$, $j=1,2,3$. These derivations act as
\begin{equation}
\partial_j a = - \imath \sum_{\bm x \in \ZM^3} x_j \, a_{\bm x}(\omega) \, u_{\bm x}
\end{equation}
and their physical representations are
\begin{equation}\label{Eq:Xder}
\pi_\omega(\partial_j a) = \imath [\pi_\omega(a),X_j].
\end{equation}
The measure ${\rm d}\mathbb P$ on $\Omega$ can be promoted to a continuous trace on $\Aa_3$ as
\begin{equation}\label{Eq:disordertrace}
\Tt (a) = \int_\Omega {\rm d}\mathbb P(\omega) \, {\rm tr}\big (a_{\bm 0} (\omega) \big ),
\end{equation}
where ${\rm tr}$ is the ordinary trace on $M_N(\CM)$. This trace is $\rho$-invariant and computes the trace per volume of covariant physical observables. Note that $\Tt$ is normalized as $\Tt(1)=N$.

\begin{remark}{\rm Since the probability measure is invariant and ergodic with respect to $\tau$, using Birkhoff’s ergodic theorem we have, $\mathbb P$-almost surely,
\begin{equation}\label{Eq:averagetrace}
\int_\Omega {\rm d}\mathbb P(\omega) \, {\rm tr}\big (a_{\bm 0} (\omega) \big)= \displaystyle{\lim_{V \to \infty}} \frac{1}{|V|} \sum_{\bm x  \in V} {\rm Tr} \big( \langle \bm x | \pi_{\omega'} (a) |\bm x \rangle \big), 
\end{equation}
where $\omega'$ is a fixed element of $\Omega$, $V$ a finite box in $\ZM^3$ and $|V|$ its volume. The right hand side of \eqref{Eq:averagetrace} is precisely the trace per volume of the physical observable $\pi_{\omega'}(a)$. The relation stated in \eqref{Eq:averagetrace} demonstrates the self-averaging property of the trace per volume, namely, that it returns a value independent of the disorder configuration $\omega'$ for a covariant observable. 
}
$\Diamond$
\end{remark}

\vspace{0.2cm}

However, our interest is not in the static setting but in cyclic deformations of the Hamiltonians resulting from adiabatic variations of the hopping functions $h_{\bm x}$. In this new setting, the hoppings become continuous maps from $\SM^1=\RM/\ZM$ to $C_N(\Omega)$. The algebra which generates such families of adiabatic Hamiltonians is simply $\bm \Aa_3=C(\SM^1,\Aa_3)$, the algebra of continuous functions from the circle to $\Aa_3$. Its elements will be indicated by bold letters. For this algebra, there is an additional 1-torus action which acts as $\bm a(t)\rightarrow \bm a(\xi t)$, and its generator provides an additional derivation $\partial_0$. The trace $\Tt$ can then be extended to $\bm \Aa_3$ as
\begin{equation}
\bm \Tt(\cdot)=\int_{\SM^1}{\rm d} t \, \Tt(\cdot),
\end{equation}
where ${\rm d} t$ denotes the unique translation invariant and normalized measure on $\SM^1$. It commutes with the action of the 4-torus generated by $\partial_j$, $j=0, 1, 2, 3$. As a consequence, we have the following general results supplied by cyclic cohomology \cite{Con}.

\begin{proposition}\label{Pro:Chern} Let $\bm p \in C^\infty(\bm \Aa_3)$ be a projection and $J \subseteq \{0, 1, 2, 3\}$ a subset of indices of even cardinality. Then the even Chern numbers
\begin{equation}
{\rm Ch}_{J}(\bm p) = \frac{(2 \pi \imath)^{\frac{|J|}{2}}}{\frac{|J| }{2} !} \sum_{\sigma \in \Ss_J} (-1)^\sigma \bm \Tt\Big ( \bm p \prod_{j \in J} \partial_{\sigma_j} \bm p \Big )
\end{equation}
are invariant against smooth deformations of $\bm p$. Similarly, let $\bm u \in C^\infty(\bm \Aa_3)$ be a unitary element and $J \subseteq \{0, 1, 2, 3\}$ a subset of odd cardinality.Then the odd Chern numbers
\begin{equation}
{\rm Ch}_J(\bm u) =  \frac{ \imath (\imath \pi)^{ \frac{|J| -1}{2}} }{|J|!!} \sum_{\sigma \in \Ss_J} (-1)^\sigma \bm \Tt\Big ( \prod_{j \in J} \bm u^{-1} \partial_{\sigma_j} \bm u \Big )
\end{equation}
are invariant against smooth deformations of $\bm u$. Above, $\Ss_J$ is the group of permutations over $J$, and $C^\infty$ indicates the sub-algebra of infinitely differentiable elements w.r.t. $\partial_j$, $j=0,\ldots,3$. 
\end{proposition}

\begin{remark}{\rm The Chern numbers can be expressed in real-space using the physical representations \eqref{Eq:Xder}, \eqref{Eq:disordertrace} and \eqref{Eq:averagetrace}. For example the even Chern numbers take the form
\begin{equation}
{\rm Ch}_{J}(\bm P_\omega) = \frac{(2 \pi \imath)^{\frac{|J|}{2}}}{\frac{|J| }{2} !} \sum_{\sigma \in \Ss_J} (-1)^\sigma \bm \Tt\Big ( \bm P_\omega \prod_{j \in J} \imath [\bm P_\omega, X_j]  \Big), \quad \bm P_\omega = \pi_\omega(\bm p). 
\end{equation}
When $J= \{0, 1 \}$ and $J=\{0, 1, 2, 3\}$, these expressions are directly connected to the real-space formulae for electric polarization \eqref{Eq:TP} and magneto-electric response, respectively \eqref{Eq:TME}.}
$\Diamond$
\end{remark}

\begin{example}{\rm For a two-dimensional (2D) disordered Hamiltonian $H_{\omega}$, its even Chern number can be expressed as
\begin{equation}\label{Eq:HallQ}
{\rm Ch}_{\{1, 2\}}(P_\omega) = - 2 \pi \imath  \Tt \Big ( P_{\omega} \big[  [ P_{\omega}, X_1],  [P_{\omega}, X_2]  \big] \Big), 
\end{equation}
where $P_{\omega} = \chi(H_{\omega})$ is the Fermi projection. The formula is quantized, independent of disorder, and stable as long as the Fermi level lies in an Anderson localized spectrum. It was used to explain the quantized plateaus in the quantum Hall effect \cite{BES}.}
$\Diamond$
\end{example}

\begin{remark}{\rm Similarly, the odd Chern numbers have physical representations
\begin{equation}
{\rm Ch}_{J}(\bm U_\omega) = \frac{ \imath ( \imath \pi)^{ \frac{|J| -1}{2}} }{|J|!!}  \sum_{\sigma \in \Ss_J} (-1)^\sigma \bm \Tt\Big ( \bm U_\omega^\ast \prod_{j \in J} \imath [\bm U_\omega, X_j]  \Big), \quad \bm U_\omega = \pi_\omega(\bm u), 
\end{equation}
which can also be connected to experimentally measurable effects \cite{PS}.}
$\Diamond$
\end{remark}

\vspace{0.2cm}

A gapped bulk Hamiltonian is a pair $(h,G)$, where $h$ is a self-adjoint element of the algebra and $G$ is a connected component of the resolvent set $\RM \setminus {\rm Spec}(h)$. We will use $G$ interchangeably as an interval or as an arbitrary point of this interval. For example, we write the projection of a gapped Hamiltonian $(h,G)$ as $p_G = \chi_{(-\infty,G]}(h)$. Throughout, $\chi_A$ represents the indicator function of the set $A$. Since $C^\ast$-algebras are stable only under the continuous functional calculus, $p_G \in \Aa_3$ only if $G \neq \emptyset$, in which case the singularity of the step function $\chi$ can be rounded without affecting the result. Similarly, we will consider gapped adiabatic Hamiltonians $(\bm h, \bm G)$ from the algebra $\bm \Aa_3$. Since ${\rm Spec}(\bm h) = \bigcup_{t \in \SM^1} {\rm Spec}\big ( h(t) \big )$, a gap $\bm G$ in the spectrum of $\bm h$ implies that all $h(t)$ are gapped at $\bm G$. In fact, $\bm G = \bigcap_{t \in \SM^1}G(t)$. Note the distinct possibility that $h(t)$ are all gapped but the gaps $G(t)$ are not aligned and as a result $\bm h$ is not gapped. The latter can be fixed by adding a family of trivial operators $\epsilon(t) \cdot 1$, to the family $h(t)$, which affects the time evolution by a trivial phase factor. Hence we can always assume that the gaps $G(t)$ are aligned such that $\bm G \neq \emptyset$. To ease the notation, we will interchangeably use $\bm G$ for the family of gaps $\{G(t)\}_{t \in  \SM^1}$, for the gap interval $\bm G = \bigcap_{t \in \SM^1}G(t)$, and for a numerical value inside the gap of $\bm h$.

\vspace{0.2cm}

Lastly, let us recall that $C^\infty$-algebras are stable under the $C^\infty$-functional calculus \cite{ProdanBook2017}. Therefore, for any $h \in C^\infty(\Aa_3)$ such as any finite range Hamiltonian, we automatically have $p_G \in C^\infty(\Aa_3)$. Similarly for the $\bm \Aa_3$ algebra.

\subsection{Time-reversal symmetric topological insulators} 
\label{Sec:TRSymmetry}

The time-reversal operation $T$ is implemented on the Hilbert space of the lattice by an anti-unitary operator. If we resolve the spin degrees of freedom, {i.e.} we make the isomorphism $\mathbb C^{2N} \simeq \mathbb C^2 \otimes \mathbb C^N$ explicit, then
\begin{equation}
T = \Big ( e^{\imath \pi \sigma_2} \otimes 1_{\mathbb C^N} \otimes 1_{\ell^2(\mathbb Z^3)} \Big )K, \quad \sigma_2 = \begin{pmatrix}  0 & -\imath \\ \imath & 0 \end{pmatrix} ,
\end{equation} 
where $K$ is the complex conjugation on $\CM^{2N} \otimes \ell^2(\ZM^3)$. It acts as
\begin{equation}
K\big ( \xi \otimes |\bm x \rangle \big ) = \bar \xi \otimes |\bm x \rangle,
\end{equation}
with the bar indicating the ordinary complex conjugation on $\CM^{2N}$.

\vspace{0.2cm}

Whenever the time-reversal symmetry is assumed, the flux matrix $\Phi$ is automatically set to zero. In such cases, the time-reversal operation acts on the Hamiltonians as
\begin{equation}
T H_\omega T^{-1} = \sum_{\bm x \in \ZM^3} \sum_{\bm y \in \ZM^3} e^{\imath \pi \sigma_2} \bar h_{\bm x}(\tau_{\bm y}\omega) e^{-\imath\pi \sigma_2} \otimes |\bm y \rangle \langle \bm y | U_{\bm x}.
\end{equation}
This drops to an anti-linear automorphism on the algebra $\Aa_3$
\begin{equation}
h  = \sum_{\bm x} h_{\bm x}(\omega) u_{\bm x} \mapsto T(h) = \sum_{\bm x \in \ZM^3} \sum_{\bm y \in \ZM^3} e^{\imath\pi \sigma_2} \bar h_{\bm x}(\tau_{\bm y}\omega) e^{-\imath \pi \sigma_2} u_{\bm x},
\end{equation}
as one can easily verify that $\pi_\omega(T(h)) = T \pi_\omega(h) T^{-1}$.  

\vspace{0.2cm}

Later, we will consider gapped adiabatic deformations which connect time-reversal symmetric (TRS) Hamiltonians. These deformations are always chosen such that
$H_\omega(2 \pi -t) = T H_\omega(t) T^\dagger$. In other words, we will consider elements $\bm h = \{h(t)\}_{t \in \SM^1} \in \bm \Aa_3$ such that $h(2\pi -t) = T\big ( h(t) \big )$. From this rule, one can see that the deformations break TRS except at two TRS points $t=0 / \pi$, where the relations $h(0) = T\big ( h(0) \big )$ and $h(\pi )= T\big ( h(\pi) \big )$ are automatically satisfied.

\subsection{Half-space and boundary physical observables}
\label{Sec:HSAlgebra}

Our investigation focuses on the physics occurring at the surfaces of the systems we just introduced. In this setting, system are restricted to half of the space, hence the Hilbert space becomes  $\widehat \Hh = \CM^{N} \otimes \ell^2(\ZM^2 \times \NM)$ and the Hamiltonians transform as
\begin{equation}\label{Eq:HalfSpaceHam}
H_\omega(t) \rightarrow \widehat H_\omega(t) = \Pi^\dagger H_\omega(t) \Pi + \widetilde H_\omega(t).
\end{equation}
Here $\Pi : \ell^2(\ZM^3) \rightarrow \ell^2(\ZM^2 \times \NM)$ is the standard partial isometry and $\widetilde H_\omega$ is a boundary Hamiltonian localized at the surface modeling some arbitrary boundary condition. On the half-space, the magnetic translation operator $\widehat U_3$ acting along the third direction is no longer unitary and instead 
\begin{equation}
\widehat U_3  \widehat U_3^\ast = I, \quad \widehat U_3^\ast  \widehat U_3 = I-P_0 , 
\end{equation}
where $P_0= \sum_{\bm x \in \ZM^2}|\bm x,0 \rangle \langle \bm x,0 |$ is the projection onto the boundary. For the static case, this reduces to the half-space algebra $\widehat \Aa_3$ defined as the universal $C^\ast$-algebra \cite{PS}
\begin{equation}
\widehat \Aa_3 = C^\ast(C_N(\Omega),\hat u_1,\hat u_2,\hat u_3),
\end{equation}
with the same commutation relations as for the bulk algebra, except for $\hat u_3^\ast \hat u_3 =1-\hat e$ where $\hat e$ a proper projection $\hat e^2=\hat e^\ast=\hat e \neq 1$. A general element $\hat a \in \Aa_3$ has the form
\begin{equation}\label{Eq:YYYYY}
\hat a =  \sum_{n, m \in \NM} \sum_{\bm x \in \ZM^2} \hat a_{\bm x, nm} \, \hat u_{\bm x} \hat u_3^n (\hat u^*_3)^m, \quad \hat a_{\bm x, nm} \in C_N(\Omega).
\end{equation}

\vspace{0.2cm}

The half-space algebra accepts the following family of canonical representations
\begin{equation}
C_N(\Omega) \ni f \rightarrow \hat \pi_\omega(f) = \sum_{\bm x \in \ZM^3} \chi_{\NM}(x_3)f(\tau_{\bm x} \omega) \otimes |\bm x \rangle \langle \bm x |, \quad \hat u_j \rightarrow \widehat S_j, \quad j=1,2,3,
\end{equation}
which generate all physical half-space Hamiltonians \eqref{Eq:HalfSpaceHam}. Inside $\widehat \Aa_3$, there is the ideal $\widetilde A_3$ generated by elements of the form $\hat a \, \hat e \, \hat a'$ for some $\hat a, \hat a' \in \widehat \Aa_3$. Such elements are localized near the boundary when represented on the physical space, hence $\widetilde \Aa_3$ can be rightfully called the boundary algebra \cite{PS}. An element $\tilde a \in \widetilde \Aa_3$ can be presented as
\begin{equation}
\tilde a =  \sum_{n, m \in \NM} \sum_{\bm x \in \ZM^2} \tilde a_{\bm x, nm} \hat u_{\bm x} \hat u_3^n \hat e (\hat u^*_3)^m, \quad \tilde a_{\bm x, nm} \in C_N(\Omega).
\end{equation}
Compared to Eq.~\eqref{Eq:YYYYY}, note the projection $\hat e$ inserted between $\hat u_3^n$ and $(\hat u^*_3)^m$.  Given a bulk Hamiltonian $h \in \Aa_3$, its corresponding half-space Hamiltonian with the Dirichlet boundary condition $\Pi H_\omega \Pi$ is generated by
\begin{equation}
\hat h_D=\sum_{\bm x, \, x_3 \geq 0}  h_{\bm x} (\omega) \hat u_1^{x_1} \hat u_2^{x_2}\hat u_3^{x_3} + \sum_{\bm x,\, x_3 < 0}  h_{\bm x} (\omega) \hat u_1^{x_1} \hat u_2^{x_2}(\hat u_3^\ast)^{|x_3|}.
\end{equation}
This generates the first term in \eqref{Eq:HalfSpaceHam}, which models an ideal situation where the boundary is cut so perfectly that the hopping matrices are not affected by the process. By adding any element $\tilde h$ from $\widetilde \Aa_3$ to $\hat h_D$, we can change Dirichlet into any other boundary condition, for example, to reflect a more realistic cutting process. The boundary element $\tilde h$ generates the second term in \eqref{Eq:HalfSpaceHam}, through the representation $\hat \pi_\omega$,
\begin{equation}
\widetilde H_\omega =  \sum_{n, m \in \NM} \sum_{\bm x \in \ZM^2} \tilde h_{\bm x, nm} (\tau_{\bm y, n}\omega) \otimes |\bm y, n \rangle \langle \bm y, n | U_{x, n-m}.
\end{equation}
In normal conditions, the distortions introduced by the cutting process are experimentally undetectable far away from the boundary, hence we will assume a cut-off at $m, n \leq \mathcal{R}$, where $ \mathcal{R}$ can be arbitrarily large but finite. 

\vspace{0.2cm}

An important relation established in \cite{PS} is the isomorphisms between $\widetilde \Aa_3$ and $M_\infty(\CM) \otimes \Aa_2$ where $\Aa_2 = C^\ast\big (C(\Omega),u_1,u_2\big )$. Moreover, the Hamiltonian $\widehat H_\omega$ inherits from \eqref{Eq:CovariantProp} the covariant property
\begin{equation}\label{Eq:HalfCovariantProp}
\widehat V_{ {\bm x}, 0}  \widehat H_\omega \widehat V_{ {\bm x}, 0}^\ast =  \widehat H_{\tau_{ {\bm x}} \omega}, 
\end{equation}
where the magnetic translations are now $\widehat V_{ {\bm x}, n} = \Pi^\dagger V_{\bm x, n} \Pi$ for $({\bm x}, n) \in \ZM^2 \times \NM$. It was also shown in \cite{PS} that a magnetic translation in the $d$-th direction $\widehat V_{ \bm 0, n}$ will effectively shift the boundary of the Hamiltonian by $n$ units. 

\vspace{0.2cm}

There is still a 2-torus action left on $\widetilde \Aa_3$ which defines the derivations $\tilde \partial_1$ and $\tilde \partial_2$. A lower semi-continuous trace, invariant to this 2-torus action, is supplied by
\begin{equation}\label{Eq:Tr10}
\widetilde \Tt(\tilde a) = \sum_{n \in \NM} \int_\Omega {\rm d}\mathbb P(\omega) \, \tilde a_{\bm 0, n}(\omega).
\end{equation}
The adiabatic families of operators for both half-space and boundary can be generated from the algebras $\hbA_3=C(\SM^1,  \widehat \Aa_3)$ and $\tbA_3=C(\SM^1,\widetilde \Aa_3)$ respectively. As in the bulk case, we can define an additional 1-torus action on $\tbA_3$ which supplies the derivation $\tilde \partial_0$. The trace \eqref{Eq:Tr10} can be canonically extended to $\tbT$ over $\tbA_3$. Chern numbers can then be defined for projections and unitary operators, hence Proposition \ref{Pro:Chern} holds also over $\tbA_3$. In general, the Chern numbers defined for projection $\tilde{\bm p}$ and unitary $\tilde{\bm u}$ from $C^\infty(\tbA_3)$ will be denoted by $\widetilde{\rm Ch}_{J} (\tilde{\bm p})$ and $\widetilde{\rm Ch}_{J} (\tilde{\bm u})$ respectively. 

\section{The engine of bulk-boundary correspondence}

In this section, we establish a diagram of exact sequences that connect the bulk and boundary algebras of adiabatically pumped systems. The diagram induces connecting maps between the $K$-theories of the bulk and boundary algebras, which ultimately allow us to analyze the surface physics of 3D adiabatically pumped insulators in Sec.~\ref{Sec:PumpHall} and Sec.~\ref{Sec:TransferHall}. We will briefly review the relevant $K$-theoretic concepts needed for this section. For more in-depth treatments, we direct the reader to the standard textbooks \cite{Bla, RLL, WO} and to \cite{PS} for an exposition within a condensed matter context.

\subsection{Elements of $K$-theory}
\label{Sec:KTh}

There are only two $K$-groups in the complex $K$-theory of a $C^\ast$-algebra $\Aa$. The first one is the $K_0(\Aa)$ group, which classifies projections
\begin{equation}
p \in M_\infty (\CM) \otimes \Aa, \quad p^2 = p^\ast=p,
\end{equation}
with respect to the von~Neumann equivalence relation
\begin{equation}\label{Eq:EquivRelation}
p \sim p' \ \ \mbox{iff}  \ \ p=vv' \ {\rm and} \ p' = v'v, 
\end{equation}
for some partial isometries $v$ and $v'$ from $M_\infty (\CM) \otimes \Aa$. Above $M_\infty (\CM)$ is the direct limit of the algebras $M_N(\CM)$ as $N \rightarrow \infty$. The equivalence class of $p$ relative to \eqref{Eq:EquivRelation} is denoted by $[p]_0$.

\begin{remark}{\rm For any projection $p$ from $M_\infty (\CM) \otimes \Aa$, there exists $N \in \NM$ such that $p \in M_N(\CM) \otimes \Aa$. This property can be seen as a generalization of the fact that any compact projection on a Hilbert space is necessarily finite rank.}
$\Diamond$
\end{remark}

\begin{remark}{\rm There are two additional equivalence relations for projections: $p \sim_u p'$ iff $p'= u p' u^\ast$ for some unitary element $u$ from $M_\infty (\CM) \otimes \Aa$, and $p \sim_h p'$ if $p$ and $p'$ can be connected by a projection homotopy in $M_\infty (\CM) \otimes \Aa$. In general, $\sim_h \Rightarrow \sim_u \Rightarrow \sim$ but when the projections come from a stable algebra $\Aa'$, {\it i.e.} $M_\infty (\CM) \otimes \Aa' \simeq \Aa'$, which is the case for $\Aa' = M_\infty (\CM) \otimes \Aa$, the three equivalence relations coincide.  
}$\Diamond$
\end{remark}

\vspace{0.2cm}

If $p \in M_N (\CM) \otimes \Aa$ and $p' \in M_M (\CM) \otimes \Aa$, then $\begin{pmatrix} p & 0 \\ 0 & p' \end{pmatrix}$ is a projection from $M_{N+M} (\CM) \otimes \Aa$ and one can define addition
\begin{equation}
[p]_0 + [p']_0 = [p \oplus p']_0 = \left [ \begin{matrix} p & 0 \\ 0 & p' \end{matrix} \right ]_0,
\end{equation}
which provides a semigroup structure on the set of equivalence classes. Then $K_0(\Aa)$ is defined as its enveloping group. 

\begin{example}\label{Ex-K0NC}{\rm The $K_0$-group of the non-commutative $d$-torus $\Aa^d_\Phi = C^\ast(u_1, u_2, \ldots, u_d)$, $u_i u_j = e^{\imath \phi_{ij}} u_j u_i$, is
\begin{equation}
K_0(\Aa^d_\Phi) = \ZM^{2^{d-1}},
\end{equation}
independently of $\Phi$ \cite{PV1980}. Its generators $[p_J]_0$ can be uniquely labeled by the subsets $J \subseteq \{1,\ldots, d\}$ of even cardinality. This assures us that, for any projection $p$ from $M_\infty (\CM) \otimes \Aa$, one has
\begin{equation}\label{Eq:K0decomp}
[p]_0 =  \sum_{|J|={\rm even}} n_J [e_J]_0, \quad n_J \in \ZM,
\end{equation} 
where the integer coefficients $n_J$ do not change as long as $p$ is deformed inside its class. In particular, two homotopic projections will display the same coefficients, hence, $\{n_J\}_{|J|={\rm even}}$ can be regarded as the complete set of topological invariants associated to $p$ or the class $[p]_0$.   
}$\Diamond$
\end{example}

\vspace{0.2cm}

The second complex $K$-group is $K_1(\Aa)$, which classifies the unitary elements
\begin{equation}
u \in M_\infty (\CM) \otimes \Aa, \quad u u^\ast = u^\ast u=1,
\end{equation}
with respect to the homotopy equivalence relation. The class of $u\in M_\infty (\CM) \otimes \Aa$ will be denoted by $[u]_1$, and the addition between these classes is defined as
\begin{equation}
[u]_1 + [v]_0 = [u \oplus v]_1 = [uv ]_1.
\end{equation} 

\begin{example}\label{Ex-K1NC}{\rm The $K_1$-group of the non-commutative $d$-torus $\Aa^d_\Phi$ is
\begin{equation}
K_1(\Aa^d_\Phi) = \ZM^{2^{d-1}},
\end{equation}
independently of $\Phi$ \cite{PV1980}. Its generators $[u_J]_1$ can be uniquely labeled by the subsets $J \subseteq \{1,\ldots, d\}$ of odd cardinality. This assures us that, for any unitary $u$ from $M_\infty (\CM) \otimes \Aa$, one has
\begin{equation}\label{Eq:K1decomp}
[u]_1 = \sum_{|J|={\rm odd}} n_J [u_J]_1, \quad n_J \in \ZM,
\end{equation} 
where the integer coefficients $n_J$ do not change as long as $u$ is deformed inside its class. In particular, two homotopic unitaries will display the same coefficients, hence, $\{n_J\}_{|J|={\rm odd}}$ can be regarded as the complete set of topological invariants associated to $u$ or the class of $[u]_1$.
}$\Diamond$
\end{example}

\vspace{0.2cm}

Lastly, we note an important result that whenever $\Omega$ is contractible, the $K$-groups of the algebras $\Aa_d = C^{\ast}(C(\Omega), u_1, u_2, \ldots, u_d)$ coincide with the $K$-groups of $\Aa^d_\Phi$ \cite{PS} 
\begin{equation}
K_{j}(\Aa_d) \simeq K_{j}(\Aa^d_\Phi), \quad  j =0, 1,
\end{equation}
and this solves the $K$-theories of the physical algebras introduced so far.

\begin{remark}{\rm For $\Aa_d$ and $J = \{1, \ldots, d \}$, the integers $n_J$ associated with the top generators $[e_J]$ and $[u_J]$ are the strong topological invariants classifying topological insulators. They have been proven to be robust in the Anderson localization regime \cite{PLB,PS_JFA_2013}. The weak topological invariants correspond to $J \subsetneq \{1, \ldots, d \}$, which are believed to be less robust. There are simple quantitative relations between $n_J$'s and the Chern numbers ${\rm Ch}_J$.}
$\Diamond$
\end{remark}
 
\subsection{Bulk-boundary correspondence for adiabatically pumped systems}

As explained in section \ref{Sec:Formalism}, our interest is in the adiabatically pumped gapped system $(\bm h, \bm G)$ from the algebra $\bm \Aa_3$. Its gapped projection $\bm p_{\bm G} = \{p_{G}(t)\}_{t \in \SM^1}$ defines a class $[\bm p_{\bm G}]_0$ in $K_0(\bm \Aa_3)$ and, together with the constant projection $\bar{\bm p}_{\bm G}=\{p_G(0)\}_{t \in \SM^1}$ in $\bm \Aa_3$, they define an element
\begin{equation}\label{Eq:SA}
[\bm p_{\bm G}]_0 - [\bar{\bm p}_{\bm G}]_0 \in K_0(\Ss \Aa_3),
\end{equation}
in the $K_0$-theory of the suspension algebra $\Ss \Aa_3$ of $\Aa_3$. The latter is the ideal of $\bm \Aa_3$ consisting of all elements $\bm a = \{a(t)\}_{t \in \SM^1}$ with $a(0)=0$. We can argue that it is the element in \eqref{Eq:SA} rather than $[\bm p_{\bm G}]_0$ that encodes the topological data of $(\bm h, \bm G)$, because in adiabatically driven experiments one maps the changes rather than the absolute values of the observables. A deeper reason for choosing the $K$-theory of $\Ss \Aa_3$ over that of $\tbA_3$ as a starting point will be stated a few lines below. 

\vspace{0.2cm}

The algebras of physical observables $\Aa_3,  \ha_3,  \widetilde \Aa_3$ and $\bm \Aa_3$ can be connected via exact sequences. Our goal is to derive the commutating diagram \eqref{Eq:KDiagram} between the $K$-theories of the bulk and the boundary, together with the associated connecting maps. One well-known short exact sequence formed by the $C^\ast$-algebras is \cite{PS}
\begin{equation}
\label{eq-BulkBoundarySeq}
\begin{diagram}
0 & \rTo  & \widetilde \Aa_3 & \rTo{ \ \ i \ \ } & \widehat \Aa_3  & \rTo{ \ \ \mathrm{ev} \ \ }  & \Aa_3 & \rTo & 0
\end{diagram}
\;,
\end{equation}
where $i$ is the inclusion map and the evaluation map is simply ${\rm ev}(\hat u_j)=u_j$. Another exact sequence that can be formed is \cite{PS}
\begin{equation}
\label{eq-SuspensionSeq}
\begin{diagram}
0 & \rTo  & \sa_3 & \rTo{ \ \ i \ \ } & C(\SM^1,\Aa_3)=\bm \Aa_3  & \rTo{ \ \ \mathrm{ev} \ \ }  & \Aa_3 & \rTo & 0
\end{diagram}
\;,
\end{equation}
where $\Ss$ always indicates the suspension of the algebra and the evaluation map is ${\rm ev}(\bm a)=a(0)$. Similar sequences exist for the suspensions of $\wh \Aa_3$ as well as $\wt \Aa_3$. They can all be combined as
\begin{equation}\label{Eq:BigDiagram}
\begin{diagram}
& & 0 & \  &  0 & \ & 0 & & \\
& & \uTo & \  &  \uTo & \ & \uTo & & \\
0 & \rTo  & \sa_3 & \rTo{ \ \ i \ \ } & \bm \Aa_3  & \rTo{\ \ \mathrm{ev} \ \ }  & \alg_3 & \rTo & 0 &
\\
& & \uTo & \  &  \uTo & \ & \uTo & & \\
0 & \rTo  & \Ss\ha_3 & \rTo{ \ \ i \ \ } & \hbA_3  & \rTo{\ \ \mathrm{ev} \ \ }  & \ha_3 & \rTo & 0 &
\\
& & \uTo & \  &  \uTo & \ & \uTo & & \\
0 & \rTo  & \Ss\widetilde \Aa_3 & \rTo{ \ \ i \ \ } & \tbA_3  & \rTo{\ \ \mathrm{ev} \ \ }  & \widetilde \Aa_3 & \rTo & 0 & \\
& & \uTo & \  &  \uTo & \ & \uTo & & \\
& & 0 & \  &  0 & \ & 0 & & 
\end{diagram}
\end{equation}

\vspace{0.2cm}

Now, recall that every exact sequence of $C^\ast$-algebras
\begin{equation}
\label{eq-BulkBoundarySeq}
\begin{diagram}
0 & \rTo  & \Jj & \rTo{ \ \ i \ \ } & \Aa  & \rTo{ \ \ \mathrm{ev} \ \ }  & \Aa/\Jj & \rTo & 0
\end{diagram}
\;
\end{equation}
induces a six-term exact sequence among the $K$-theories \cite{RLL, WO}
\begin{equation}\label{SixTermDiagramAlg}
\begin{diagram}
& K_0(\Jj) & \rTo{ \ \ i_\ast \ \ } & K_0(\Aa)  & \rTo{\ \ {\rm ev}_\ast \ \ } & K_0(\Aa/\Jj) &\\
& \uTo{\rm Ind}& \  &  \ & \ & \dTo{\rm Exp} & \\
& K_1(\Aa/\Jj)  & \lTo{\ \ {\rm ev}_\ast \ \ } & K_1(\Aa) & \lTo{\ \ i_\ast \ \ } & K_1(\Jj) &
\end{diagram}
\end{equation}
The six-term exact sequence is the central ingredient in deriving the bulk-boundary correspondence of topological insulators \cite{KRS,PS}. Let us point out that for the exact sequence \eqref{eq-SuspensionSeq}, the connecting maps are trivial and the top part of the exact 6-term sequence reduces to
\begin{equation}\label{Eq:YY1}
\begin{diagram}
0 & \rTo & K_0(\sa_3) & \rTo{ \ \ i_\ast \ \ } & K_0(\bm \Aa_3)  & \rTo{ \ \ {\rm ev}_\ast \ \ } & K_0(\Aa_3) & \rTo & 0.
\end{diagram}
\end{equation}
This tells us that $K_0(\bm \Aa_3) = K_0(\sa_3) \oplus K_0(\Aa_3)$ and this decomposition is canonically implemented by $[\bm p]_0= \big ( [\bm p]_0 - [\bar{\bm p}]_0 \big ) + [\bar{\bm p}]_0$, with $\bar{\bm p}=\{p(0)\}_{t \in \SM^1}$. In particular, one can define a surjective homomorphism
\begin{equation}
i_\ast^{-1} : K_0(\bm \Aa_3) \rightarrow K_0(\sa_3), \quad i_\ast^{-1}\big ([\bm p]_0\big ) = [\bm p]_0 - [\bar{\bm p}]_0,
\end{equation}
which is left-inverse to $i_\ast$. This further justifies the use of \eqref{Eq:SA} in our calculations. Let us recall that $K_0(\sa_3)$ is isomorphic to $K_1(\Aa_3)$, with the isomorphism implemented by the $\theta$ map described, for example, in  \cite[Theorem.~10.1.3]{RLL} or \cite[Theorem.~7.2.5]{WO}, and that $K_1(\sa_3)$ is isomorphic to $K_0(\Aa_3)$, with the isomorphism implemented by the $\beta$ map described, for example, in  \cite[Sec.~11.1.1]{RLL} or \cite[Sec.~9.1.2]{WO}. The $\theta$ and $\beta$ maps are at the heart of the Bott periodicity theorem.

\vspace{0.2cm}

We now apply \eqref{SixTermDiagramAlg} to the vertical exact sequences appearing at the left and at the right in \eqref{Eq:BigDiagram} as well as to the top horizontal one and, by using the $\theta$ and $\beta$ maps, we produce the following commutative diagram
\begin{equation}\label{Eq:KDiagram}
\begin{diagram}
K_0(\bm \Aa_3)  & \rTo{\ \  i_\ast^{-1} \ \ } & K_0(\sa_3)   &  \rTo{\ \ \theta^{-1} \ \ } & K_1(\alg_3) &\\
& & \dTo{\rm Exp}& \  &   \dTo{\rm Ind} &   \\
& & K_1(\Ss \widetilde \Aa_3) &  \rTo{\ \ \beta^{-1}\ \ } & K_0(\widetilde \Aa_3) &,
\end{diagram}
\end{equation}
which will be our main tool to derive the bulk-boundary correspondence. Let us state that the main reason for starting the analysis from $\Ss \widetilde \Aa_3$ instead of $C(\SM^1,\widetilde \Aa_3)$ is to put these $\theta$ and $\beta$ standard maps to work. This reflects a general principle about adiabatically driven systems, which was discovered in \cite{PS}[Sec.~4.3.4], pointing to the suspension algebra as the natural environment to analyze such cycles.

\vspace{0.2cm}

According to \eqref{Eq:KDiagram}, the topological data of the bulk Hamiltonian $(\bm h, \bm G)$ may be stored at the boundary in an element from $K_1(\Ss \widetilde \Aa_3)$ or $K_0(\widetilde \Aa_3)$. As such, there are two alternative routes to explore the boundary physics: 1) through ${\rm Exp}$ and 2) through $\beta^{-1} \circ {\rm Exp}$ or ${\rm Ind} \circ \theta^{-1}$. Via the first path, we land in $K_1(\Ss \widetilde \Aa_3)$. As we shall see in section~\ref{Sec:PumpHall}, this allows us to characterize the flow of the energy spectrum during an adiabatic cycle. Via the second route, we land in $K_0(\widetilde \Aa_3)$. As we shall see in section~\ref{Sec:TransferHall}, this allows us to characterize the cumulative changes after one full adiabatic cycle. Furthermore, the explicit computation of the connecting maps will reveal specific experimental settings where these changes become interesting. Let us conclude with the remark that navigating the second route via $\beta^{-1} \circ {\rm Exp}$ or ${\rm Ind} \circ \theta^{-1}$ should make no difference, simply because the diagram \eqref{Eq:KDiagram} is commutative. However, to our knowledge, the inverse map $\beta^{-1}$ is not known explicitly while $\theta^{-1}$ can be found for example in \cite[pg.~111]{PS}. As such, we will navigate the second route using ${\rm Ind} \circ \theta^{-1}$.

\section{Pumping of surface Hall states}
\label{Sec:PumpHall}

Here we use the exponential map to establish the bulk-boundary correspondence for the spectral flow at the surface of a 3D insulator under an adiabatic cycle. Using well known relations between various topological invariants, we predict two interesting physical surface phenomena.

\subsection{Surface spectral flow and bulk-boundary correspondence}
\label{Sec:SpecFlow}

The surface spectral flow refers to the evolution of ${\rm Spec}\big (\hat{h}(t)\big ) \cap \bm G$ under the adiabatic deformation and it is a measure of charge pumping. A general definition and quantitative characterization of the spectral flow can be found in \cite{CPRS2006}, within the framework of Brewer index on von Neumann algebras. We provide a definition of net surface  spectral flow based on a more physical picture.

\begin{definition}[Surface spectral flow] Let $(\bm h,\bm G)$ be a bulk gapped Hamiltonian and $\hat{\bm h}$ its half-space restriction with an arbitrary boundary condition. Let $\bm{\widehat H}_\omega^A$ be a finite-area approximation of $\hat \pi_\omega(\hat{\bm h})$, obtained by confining the coordinates in the first and second directions to a finite area $A$. It is immediate to see that the spectra ${\rm Spec}\big(\widehat H_\omega^A(t)\big ) \cap G(t)$ consist of discrete eigenvalues. Let $E:\SM^1 \rightarrow \RM$ be continuous function such that $E(t) \in G(t)$ for all $t \in \SM^1$, but otherwise arbitrary. Then the net spectral flow per adiabatic cycle is defined as
\begin{equation}\label{Eq:SF}
{\rm Sf}\big ( \hat{\bm h}, \bm G\big ) = \lim_{A \rightarrow \infty} \int_\Omega {\rm d}\mathbb P(\omega) \ \frac{N_{\omega,+}^A - N_{\omega,-}^A}{A},
\end{equation}
where $N_{\omega,\pm}^A$ count the number of times an eigenvalue of $\widehat H_\omega^A(t)$ crosses the curve $E(t)$ from below/above, respectively, during the full cycle $t \in \SM^1$. 
\end{definition}

\begin{remark}{\rm The number defined in \eqref{Eq:SF} can be shown to be independent of the choice of $E(t)$ or of the boundary condition used to confine the coordinates in area $A$ \cite{CPRS2006}, hence \eqref{Eq:SF} is a true characteristic of the gap $\bm G$. If one starts the adiabatic cycle from a trivial TRS phase with gapped surface spectrum, then it is immediate to see that the spectral flow \eqref{Eq:SF} measures the number of electrons per unit area exchanged by the valence and conduction bands after one full adiabatic cycle.}
$\Diamond$ 
\end{remark}

\vspace{0.2cm}

We now compute the exponential map ${\rm Exp}$ in \eqref{Eq:KDiagram} and establish the bulk-boundary correspondence for ${\rm Sf}(\hat{\bm h},\bm G)$. Henceforth, let $(\bm h,\bm G)$ be a gapped Hamiltonian from $\bm \Aa_3$. From \cite{PS}[Proposition.~4.3.1], we have
\begin{equation}
{\rm Exp}\big ( [\bm p_{\bm G}]_0 - [\bar{\bm p}_{\bm G}]_0\big ) = \big [\tilde{\bm u}_{\bm G} \, \bar{\bm u}_{\bm G}^{-1}]_1, 
\end{equation}
with the unitary elements defined by 
\begin{equation}
\tilde {\bm u}_{\bm G}= \exp\big (2 \pi \imath f(\hat{\bm h}) \big ), \quad \bar {\bm u}_{\bm G}= \exp\big (2 \pi \imath f(\bar{\bm h}) \big ),
\end{equation}
where $f:\RM \rightarrow \RM$ is a non-decreasing smooth function taking values 0/1 below/above $\bm G$ and $\bar{\bm h}$ is the family of constant Hamiltonians $\bar{h}(t)=\hat{h}(0)$. The function $f$ is arbitrary except for the constraints we just mentioned. In particular, its entire variation can be concentrated around any point in $\bm G$. One can convince oneself that $\tilde{\bm u}_{\bm G} - 1$ and $\bar{\bm u}_{\bm G} - 1$ belong to $\tbA_3$, and in fact to $C^\infty(\tbA_3)$. While $\bm p_G$ encodes the topological data of the bulk, $\tilde{\bm u}_G$ encodes the topological data of the boundary.

\vspace{0.2cm}

There is abundant information about the relations between these elements. Below we select several results from the literature that facilitate Proposition~\ref{Pro:SF}, which is one of our important tools for the analysis of spectrum during adiabatic cycles.

\begin{proposition}\label{Pro:SpecFlow} The following statements hold
\begin{enumerate}[{\rm i)}]
\item Form {\rm \cite{PS}[Theorem.~5.5.3]}:
\begin{equation}
{\rm Ch}_{\{0,3\}}(\bm p_{\bm G}) = \widetilde{\rm Ch}_{\{0\}}(\tilde{\bm u}_{\bm G}).
\end{equation}
\item From {\rm \cite{CPRS2006}[Theorem.~4.2]}:
\begin{equation}
 \widetilde{\rm Ch}_{\{0\}}(\tilde{\bm u}_{\bm G}) = - {\rm Sf}\big ( \hat{\bm h},\bm G\big ).
 \end{equation}
\item From {\rm \cite{PS}[Theorem.~5.6.3 \& Corollary.~5.6.4]}:
\begin{equation}
\partial_{\phi_{12}} {\rm Ch}_{\{0,3\}}(\bm p_{\bm G}) = {\rm Ch}_{\{0,1,2,3\}}(\bm p_{\bm G}), 
\end{equation}
where $\phi_{12}$ is the entry at position {\rm 1-2} in the flux matrix $\Phi$.
 \end{enumerate}
 \end{proposition}
 
 \begin{remark}{\rm i) is a standard bulk boundary-correspondence, connecting a bulk (weak) invariant to its corresponding surface invariant; ii) connects a measurable quantity, the spectral flow, with a mathematical object, $\widetilde{\rm Ch}_{\{0\}}(\tilde{\bm u}_{\bm G})$; iii) is a generalized Streda formula.
 }$\Diamond$
 \end{remark} 

\begin{proposition}\label{Pro:SF} The surface spectral flow manifests the following bulk-boundary correspondence 
\begin{equation}\label{Eq:SFvsME}
\partial_{\phi_{12}} {\rm Sf}\big ( \hat{\bm h}, \bm G\big ) = -  \Delta \alpha_{\rm ME}.
 \end{equation}
 \end{proposition}

 \proof From i) and ii) of Proposition~\ref{Pro:SF}
 \begin{equation}\label{Eq:Use1}
 {\rm Sf}\big ( \hat{\bm h}, \bm G\big ) = - {\rm Ch}_{\{0,3\}}.
 \end{equation}
 Taking the derivative w.r.t. the magnetic flux $\phi_{12}$ and using the generalized Streda formula stated at iii) of Proposition~\ref{Pro:SF},
  \begin{equation}\label{Eq:Use2}
\partial_{\phi_{12}} {\rm Sf}\big ( \hat{\bm h}, \bm G\big ) = - {\rm Ch}_{\{0,1,2,3\}}(\bm p_{\bm G}).
 \end{equation}
Recalling our introductory remarks about the magneto-electric response, we have $\Delta \alpha_{\rm ME} = {\rm Ch}_{\{0,1,2,3\}}(\bm p_{\bm G})$ and the statement follows. \qed

\vspace{0.2cm}

Proposition~\ref{Pro:SF} is a bulk-boundary correspondence between physically measurable quantities. The magneto-electric response coefficient is a bulk observable that can be measured in a configuration without a surface. The spectral flow is a surface observable and its measurement involves a surface experiment. Yet, Eq.~\eqref{Eq:SFvsME} relates these two independent measurements. Furthermore, since ${\rm Sf}(\hat{\bm h},\bm G)$ quantifies the flow of the surface spectrum, we can now make an important spectral statement. Assume a non-trivial change in the bulk magneto-electric response coefficient,
\begin{equation}
\Delta \alpha_{\rm ME} = {\rm Ch}_{\{0,1,2,3\}}(\bm p_{\bm G}) \neq 0.
\end{equation} 
Then identity \eqref{Eq:Use1} implies that $\hat{\bm h}$ cannot be gapped at $\bm G$. Otherwise, no spectrum can flow across the gap, hence, ${\rm Sf}(\hat{\bm h},\bm G)$ should be identically zero and \eqref{Eq:Use1} will be violated. 

\subsection{Physical picture}
\label{Sec:Phys1}

\begin{figure}
 \center
\includegraphics[width=0.6\linewidth]{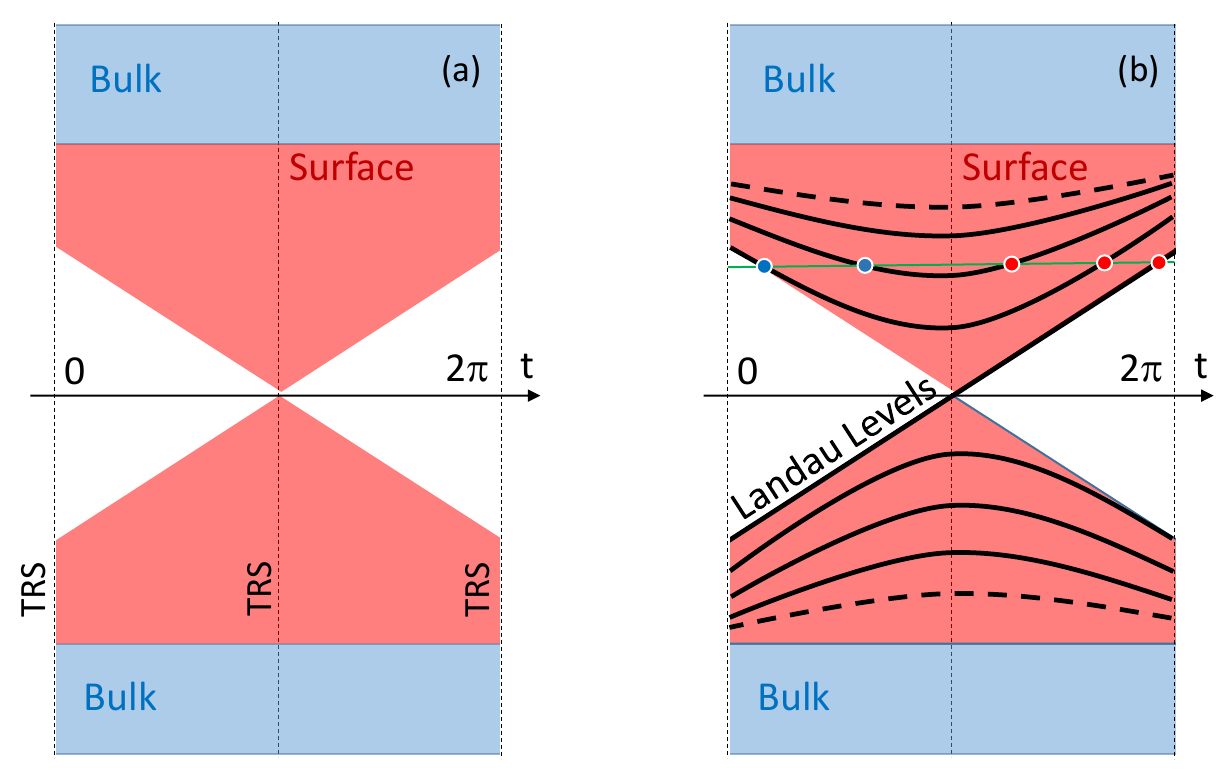}
\caption{\small {\bf Surface spectral flow for $\Delta \alpha_{\rm ME}=-1$.} (a) A typical spectral flow when the adiabatic cycle connects a trivial, at $t=0$, and a TRS-stabilized topological insulator, at $t=\pi$. The red shaded region represent the continuum part of the boundary spectrum and a Dirac cone is visible at $t=\pi$. (b) When a small magnetic field is applied to the surface of the insulator, the continuum boundary spectrum, shown again in red color, splits into Landau levels, whose evolutions in time are represented by the solid lines. We call them Landau bands and all the Landau bands in panel (b) are quadratic except for one which is chiral. A counting of the Landau level crossings over a fixed energy level (see the green line) is illustrated, with the outcome $\Mm_-=2$ (blue dots) and $\Mm_+=3$ (red dots).}
\label{Fig:SpecFlow1}
\end{figure}

A detailed picture of the spectral flow can be obtained when the magnetic flux $\phi_{12} \neq 0$ and the disorder is weak. In this case, the boundary spectrum of $\hat{ h}(t)$ consists of Landau levels, a well-known fact coming from both simulations and experiments \cite{Cheng2010, Hanaguri2010}. Since we are dealing with crystals, the Landau levels have finite widths and complicated fractal structure. However, the Landau levels can still be rigorously defined as isolated islands of (fractal) spectrum whose corresponding spectral projectors belong to the $K_0$-class $\pm[\tilde e_{\{1,2\}}]_0$ of the boundary algebra, in the notation of section~\ref{Sec:KTh}. Equivalently, these Landau levels carry a Chern number $\pm 1$. In Fig.~\ref{Fig:SpecFlow1}(b), we consider the case of weak magnetic field when the width of Landau levels can be ignored (see {\it e.g.} Fig.~7.1 in \cite{ProdanBook2017}). The flow of the Landau levels can be drawn as curves and we shall call them Landau bands. This regime of small $\phi_{12}$ will be automatically assumed every time when Landau bands are mentioned. 

\vspace{0.2cm}
 
\vspace{0.2cm}

The results of the previous section can be reformulated in terms of Landau bands as follows:
 
 \begin{corollary}\label{Cor:Landau} Assume that the boundary spectra of $\hat h(t)$ can be resolved in Landau levels when $\phi_{12}$ is turned on. Under the adiabatic cycle, the Landau levels flow as in Fig.~\ref{Fig:SpecFlow1}(b) when a gapped initial $\hat{h} (0)$ is assumed. Specifically, there must be a number $\Nn_+$ of Landau levels that emerge from the spectrum $Spec_-(\bm h)$ below $\bm G$ and dive into the spectrum $Spec_+(\bm h)$ above $\bm G$, as well as a number $\Nn_-$ of Landau levels that do the opposite, and the two numbers are related by
 \begin{equation}
 \Nn_+ - \Nn_- = - {\rm Ch}_{\{0,1,2,3\}}(\bm p_G)=- \Delta \alpha_{\rm ME}.
 \end{equation}
 \end{corollary} 
 
 \proof We know that boundary spectrum consists of Landau levels in the presence of a magnetic field perpendicular to the surface. Each Landau level carries an electron density equal to $\phi_{12}$ because $\widetilde{\Tt}(\tilde e_{\{1,2\}})=\phi_{12}$. The amount of spectrum per unit area that crosses an arbitrary but fixed energy level during the adiabatic cycle is $(\Mm_+ - \Mm_-) \phi_{12}$, where $\Mm_\pm$ counts the number of times a Landau band crosses that energy level from below/above during the cycle. Since the quadratic Landau bands, {\it i.e.} those who originate and return to the same part of the bulk spectrum, do not contribute to the difference $\Mm_+ - \Mm_-$, the statement follows.\qed
 
\vspace{0.2cm}

The $\ZM_2$ topological invariant classifying the TRS-stabilized topological insulators can be defined as $\Delta \alpha_{\rm ME} \, {\rm mod}\, 2$, where the change $\Delta$ is with respect to any closed, TRS-respecting adiabatic path that connects the insulator with a standard trivial reference \cite{LP}. Hence, when the physical adiabatic cycle considered in our work connects a trivial insulator ($t=0$) and a topological TRS-stabilized insulator ($t=\pi$), ${\rm Ch}_{\{0,1,2,3,4\}}(\bm p_{\bm G})$ is necessarily an odd integer. Fig.~\ref{Fig:SpecFlow1} illustrates a possible spectral flow for such a case. It is tailored to the simplest case possible, where the gap of the surface spectrum closes only for the TRS-stabilized topological phase. Furthermore, it is assumed in Fig.~\ref{Fig:SpecFlow1} that the gapless topological surface spectrum displays a single Dirac singularity and  ${\rm Ch}_{\{0,1,2,3,4\}}(\bm p_{\bm G})=-1$. Then, in the absence of a magnetic field, the flow of the spectrum should look as in Fig.~\ref{Fig:SpecFlow1}(a) and, when a small magnetic field is turned on perpendicular to the surface, the flow should look as in Fig.~\ref{Fig:SpecFlow1}(b). The first thing to notice in this figure is the distinct trajectory of the middle Landau level and how it connects to the Landau bands immediately below and above. We refer to such Landau bands as a chiral Landau bands. In fact, all Landau bands shown in Fig.~\ref{Fig:SpecFlow1}(b) are connected into one single Landau band that flows from ${\rm Spec}_-(\bm h)$ to ${\rm Spec}_+(\bm h)$. For the reader's convenience, we show in Fig.~\ref{Fig:SpecFlow1}(b) an energy level and a count of $\Mm_\pm$. As one can see, no matter where we place this energy level, the difference $\Mm_+ - \Mm_-$ will always be 1 in this particular case.

\vspace{0.2cm}

The spectral flow shown in Fig.~\ref{Fig:SpecFlow1}(b) can also be easily understood starting from the physics of 2D Dirac systems in a uniform magnetic field, which will be treated in more detail in section~\ref{Sec:halfIQH}. Recall that in Fig.~\ref{Fig:SpecFlow1} we assume that, at $t=\pi$ and in the absence of a magnetic field, the TRS-stabilized topological insulator displays a single Dirac cone on the surface. Then, when the magnetic field is turned on, the chiral Landau level mentioned above is precisely the Landau level trapped by the tip of the Dirac cone (see section~\ref{Sec:halfIQH}). Since the tip traps one and only one Landau level, starting from the spectral flow shown in Fig.~\ref{Fig:SpecFlow1}(a) and resolving it into Landau bands, one can readily conclude that the only allowed trajectories of the Landau levels are the ones depicted in Fig.~\ref{Fig:SpecFlow1}(b).

\begin{figure}
\center
\includegraphics[width=0.8\linewidth]{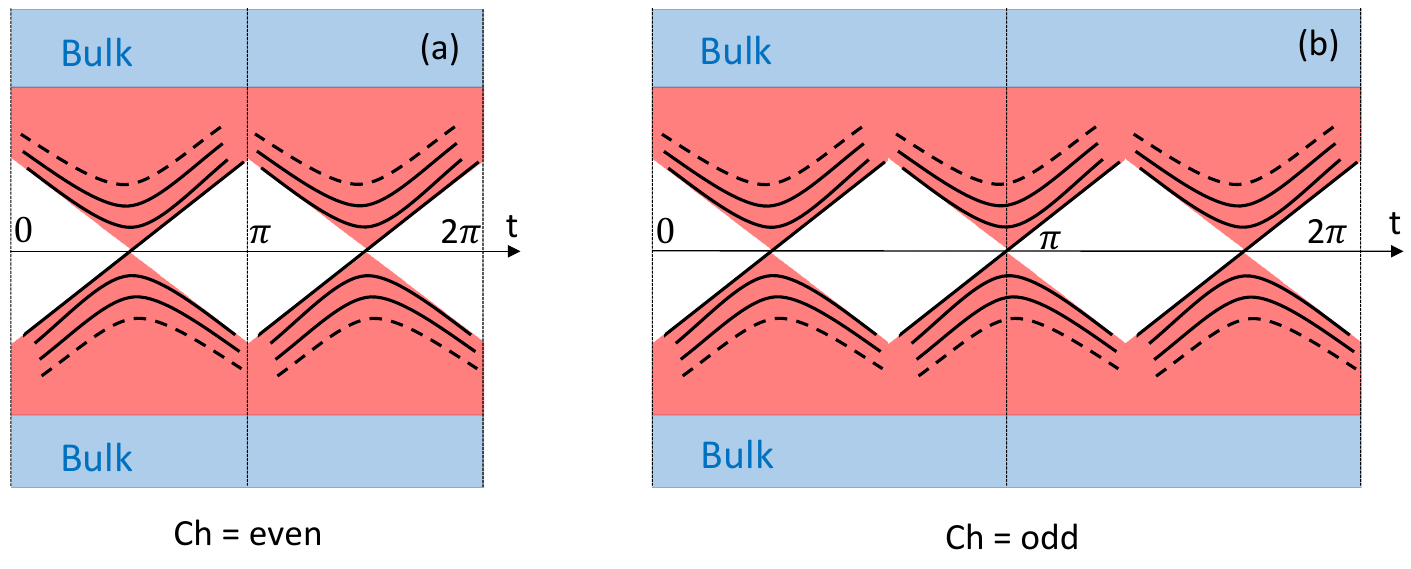}
\caption{\small {\bf Spectral flow for higher values of $\Delta \alpha_{\rm ME}$.} The solid lines represent the orbits of the Landau levels while the red shaded region represent the continuum boundary spectrum in which the Landau levels collapse when $\phi_{12} \rightarrow 0$. When $\phi_{12}=0$, the topological gap-closings are symmetric relative to the TRS-symmetric points and a qualitative difference emerges between even (left) and odd (right) values of $\Delta \alpha_{\rm ME}$.}
\label{Fig:SpecFlow2}
\end{figure}

\vspace{0.2cm}

In general, the numbers $\Nn_\pm$ of chiral Landau bands must obey the constraint $\Nn_+ - \Nn_-=- {\rm Ch}_{\{0,1,2,3\}}(\bm p_{\bm G})$. Let us restrict the discussion to the stable case when either $\Nn_+$ or $\Nn_-$ is zero, in which case the flow of the Landau levels must look as in Fig.~\ref{Fig:SpecFlow2}. Note that this stable configuration can always be met via small perturbations. When the magnetic field is turned off, the Landau bands dissolve back into boundary essential spectrum. Hence the surface spectral flow in the absence of a magnetic field must be as indicated by the red-shaded regions in Fig.~\ref{Fig:SpecFlow2}. The conclusion is that the boundary spectrum ${\rm Spec}\big (\hat h(t) \big ) \cap \bm G$ must close along the adiabatic cycle at least a number of times equal to $\Delta \alpha_{\rm ME}$. Note that we cannot rule out accidental gap-closings along the adiabatic path. If the bulk-gap is large, then such events are very rare and can be eliminated by small perturbations. The gap-closings that cannot be removed by small perturbations will be called topological from now on. All topological gap-closings along the adiabatic cycle occur through Weyl singularities in the $(t,k_1,k_2)$-space and $\widetilde{\rm Ch}_{\{0,1,2\}}(\hat{\bm u}_G)$ measures the chiralities of these Weyl points (see Example~5.3.3 in \cite{PS}). At the moment $t$ when a topological gap closes, one will observe a Dirac cone at the surface of the insulator. In the presence of a weak magnetic field, the chiral Landau bands depicted in Fig.~\ref{Fig:SpecFlow2} correspond to the Landau levels trapped by the tips of these Dirac cones.

\vspace{0.2cm}

If the magnetic flux $\phi_{12}$ is turned off and then turned on with opposite sign, the flow of the Landau levels will be the mirror reflection of the flow seen in Fig.~\ref{Fig:SpecFlow2}, relative to the TRS points $t=0$ and $t=\pi$. As such, at $\phi_{12}=0$, the topological gap-closings must occur symmetric relative to the TRS points. From this observation, one can see the major difference between odd and even bulk Chern numbers, which is quite apparent in Fig.~\ref{Fig:SpecFlow2}. If the value of ${\rm Ch}_{\{0,1,2,3\}}(\bm p_{\bm G})$ is odd, then there is necessarily one topological gap-closing at one of the TRS points which we always choose to be $\pi$. This gap-closing cannot be lifted by any continuous deformations of the insulator and defines the TRS-stabilized topological phase in 3D. Hence, our analysis supplies yet another way to understand why this phase has un-gapped boundary spectrum. Note that the analysis is completely independent of boundary conditions as the only assumption is that the adiabatic cycle respects the TRS operation.

\subsection{Experimental predictions I}

Suppose that one drives a TRS-stabilized topological insulator in an adiabatic cycle such that there is a net quantized change in the bulk magneto-electric response. Then the experimentalist will observe the following surface phenomena:
\begin{itemize}

\item Recall that the net spectral flow is independent of the reference energy line $E(t)$, as long as this line takes values inside $\bm G$, is continuous and closes into itself. By placing the line infinitesimally close to the bottom/top of the gap $\bm G$, we can see that the spectral flow implies that electrons are pumped from the valence to the conduction band at a rate of $\phi_{12}\Delta \alpha_{\rm ME}$ per cycle. 

\item Instead of a halved sample, suppose one has a thick slab such that the top and bottom surfaces are decoupled. Yet a Hall measurement can be simultaneously performed on the two surfaces. Under an adiabatic cycle, the flow of the Landau levels on the top and bottom surfaces occur in opposite directions but all the other aspects remain as described above. If a Hall measurement is performed with a weak magnetic field $\phi_{12}$ on the system along the adiabatic cycle and the Fermi level is held constant, then one will observe generically a number of $2 \Delta \alpha_{\rm ME}$ plateau-plateau phase transitions. However, the net variation of the Hall conductance of the slab per cycle is zero. For more on the actual values of the Hall conductance during such processes, see next section.

\end{itemize}

\section{Surface Hall effect}

In this section, we review the qualitative arguments as well as the experimental evidence for the half-integer quantum Hall effect at the surface of a TRS-stabilized topological insulator. We then use Corollary~\ref{Cor:Landau} to give a full account of these effects in settings relevant to the experiments. 

\subsection{Half-integer quantum Hall effect at the surface of TRS topological insulators} 
\label{Sec:halfIQH}

Let us start by stating that the Hall conductance of a semi-infinite sample is not a well-defined physical observable. There is no self-averaging principle for this quantity, hence its value will fluctuate from one surface or disorder configuration to another.  This is primarily because the bulk and the surface responses are mixed together. However, there is an idealized limit where the Hall response of the surface can be separated, and this happens when the bulk gap is very large, practically infinite. In this limit and in the absence of disorder, the quantum dynamics of the surface states of an ideally halved sample is generated by the 2D Dirac operator
\begin{equation}\label{Eq:BDirac}
D_{\vec B} = (\partial_x -\imath e A_x) \, \sigma_x + (\partial_y - \imath e A_y) \, \sigma_y,
\end{equation}
in some properly chosen energy units. We pin the origin of the energy axis at the Dirac singularity of the surface band spectrum. In \eqref{Eq:BDirac}, a magnetic field $\vec B$ perpendicular to the surface, defined by the vector potential $\vec A$, was assumed. 

\vspace{0.2cm}

The eigenvalue problem for $D_{\vec B}$ can be solved analytically and the spectrum consists of Landau levels located at the energies $E_n = {\rm sgn}(n) \sqrt{2 e B |n|}$ where $n \in \ZM$. The Landau level corresponding to $n=0$ was referred in the previous section as the Landau level pinned by the tip of the Dirac cone. Each Landau level carries a first Chern number which equals to one, hence the surface Hall conductance will jump by one every time the Fermi level crosses a Landau level from below. This information alone is not enough to define the actual value of the Hall conductance because one also needs a reference point. For Hamiltonians with bounded spectra, the reference point corresponds to a Fermi energy situated below or above the spectra, where it is known that the Hall conductance is zero. But the spectrum of the magnetic Dirac operator is unbounded below and above, hence such reference point is missing. Pinpointing the net value of the surface Hall conductance rests on the chiral symmetry of the Hamiltonian. Since $\sigma_z D_{\vec B} \sigma_z = - D_{\vec B}$, the spectrum is symmetric relative to the origin and if we place the Fermi energy above $E_0$, the Fermi projections satisfy the following relation: $P_{-E_F} = I - \sigma_z P_{E_F} \sigma_z$. With the assumption that the Hall conductance $\tilde \sigma_H$ of the surface takes finite values, supported by the explicit computations in \cite{WatanabePRB}, we can then conclude that $\tilde \sigma_H(E_F) = - \tilde \sigma_H(-E_F)$ by just examining the formula \eqref{Eq:HallQ}. Now, if we place $E_F$ immediately above $E_0$, then $-E_F$ and $E_F$ are separated by just one Landau level, the one corresponding to $E_0$.  As a result, $\tilde\sigma_H(E_F)-\tilde \sigma_H(-E_F)=-1$. This allows one to conclude that $\tilde \sigma_H(\mp E_F) = \pm \frac{1}{2}$. To summarize, by using the chiral symmetry together with the assumption that $\tilde \sigma_H$ is finite, we have found two reference values of the Hall conductance that are half-integer. It is now clear that $\tilde\sigma_H$ is quantized to half-integers in general. This is also the case if a generic odd number of Dirac singularities are present in the surface band spectrum, as it could happen for a TRS-stabilized topological insulator. 

\vspace{0.2cm}

The actual Hall experiments are performed on slabs of TRS-stabilized insulators, in which case the Hall responses of the top and bottom surfaces are mixed. Nevertheless, the half-integer quantization discussed above has a dramatic experimental signature, namely, the odd-integer quantized Hall response of the combined top and bottom surfaces, an effect which has been observed in laboratories just a few years ago \cite{Zhang2014, Yoshimi2015}. In general, the Landau levels of the top and bottom surfaces need not to be aligned, but any misalignment can be corrected using gating potentials. When such alignment was achieved in the laboratory, the net Hall conductance of the slab-sample displayed {\it only odd-integer} plateaus under the variation of the magnetic field. It was argued that this is possible only if the Hall response of each surface was quantized to half-integers.

\begin{figure}
\center
\includegraphics[width=0.4\linewidth]{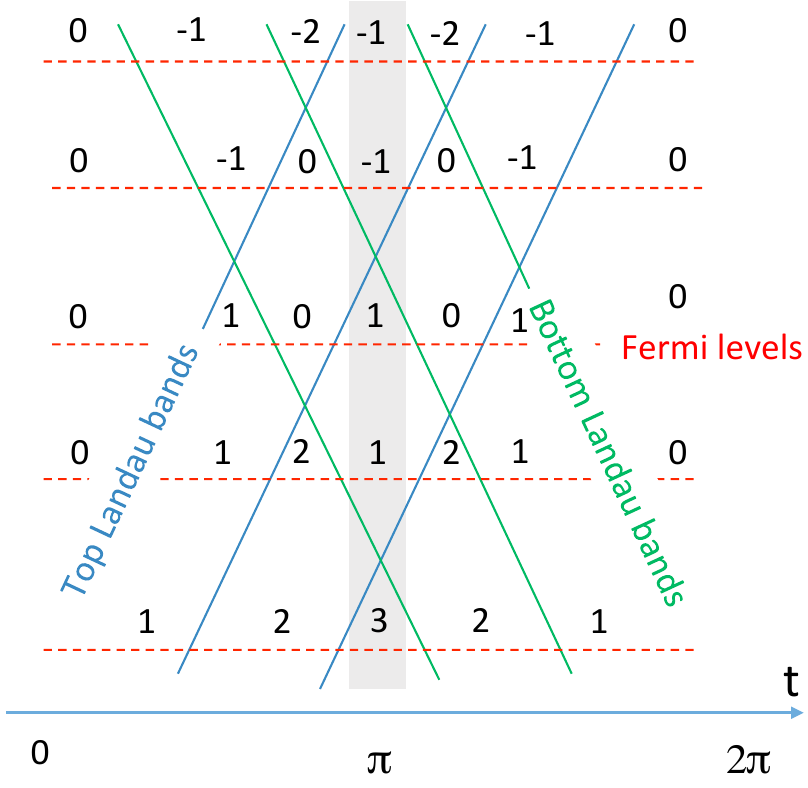}
\caption{\small {\bf The flow of Landau levels for a slab geometry.} This example corresponds to the case when $\Delta \alpha_{\rm ME} = -3$ per adiabatic cycle (see the second case in Fig.~\ref{Fig:SpecFlow2}). The blue and green solid lines represent flows of the Landau levels at the top and bottom surfaces, respectively. The integer values in the grey shaded area denote the net Hall conductance of the topological insulating slab at the TRS point $\pi$, for different Fermi levels (red dotted lines). The horizontal sequences of numbers represent the values of the slab's Hall conductance during the adiabatic cycle, for different Fermi levels.}
\label{Fig:HalfIQHE}
\end{figure}

\vspace{0.2cm} 

The qualitative arguments presented in the physics literature, including the one we sketched above, cannot account for the behavior of the Hall response under a continuous deformation of $D_{\vec B}$. There is no topological invariant associated to such half-integer quantized Hall conductivity, hence the faith of this half-integer quantization cannot be assessed when the bulk states are reintroduced in the models. This latter aspect is very difficult, if not impossible, to probe numerically. However, the experimental findings of odd-integer plateaus can be easily understood without invoking any half-integer Hall quantization if the spectral flow derived in section~\ref{Sec:SpecFlow} is used, particularly Corollary~\ref{Cor:Landau}. In Fig.~\ref{Fig:HalfIQHE}, we illustrate the flow of the Landau levels for a slab-sample subjected to a perpendicular magnetic field that penetrates the entire slab. For the sake of the argument, we choose the case when $\Delta \alpha_{\rm ME} = -3$ per adiabatic cycle but the main conclusion will be the same for any odd $\Delta \alpha_{\rm ME}$. In such cases, there is an odd number of top Landau levels flowing through the central region of the bulk gap, as seen already in Fig.~\ref{Fig:SpecFlow2}. The flows of the Landau levels on the bottom surface are in opposite direction. Every time the Fermi level crosses an upward/downward moving Landau level, the Hall conductance changes by $\pm 1$. Assume now that we start a slow deformation process from a trivial insulator at $t=0$, which by definition has $\tilde \sigma_H=0$. If the the top/bottom Landau levels are aligned at $t=\pi$, then, as illustrated in Fig.~\ref{Fig:HalfIQHE}, $\sigma_H$ is necessarily an odd integer regardless of the position of the Fermi level.

\subsection{Anomalous Hall effect and axion insulators}

\begin{figure}
\center
\includegraphics[width=0.8\linewidth]{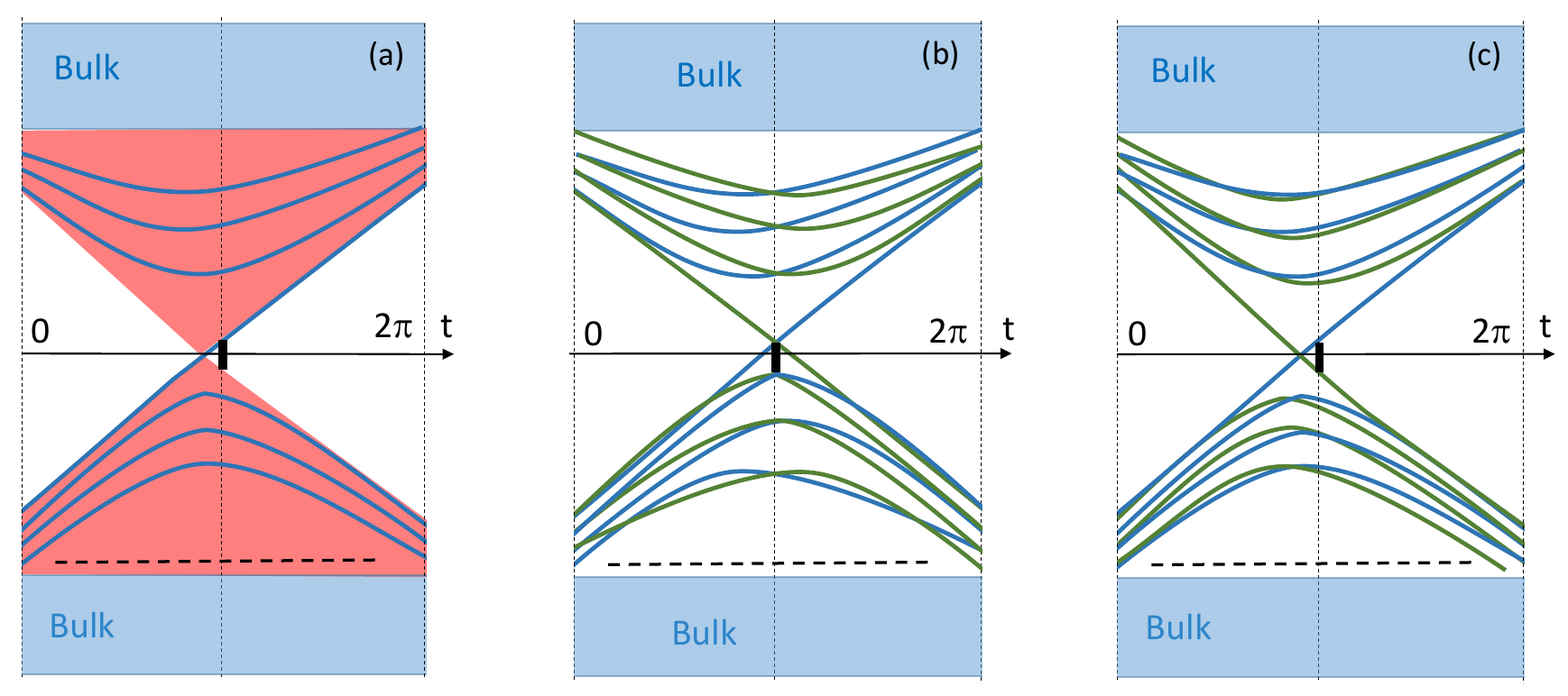}
\caption{\small {\bf The surface spectral flow when ferromagnetic films are inserted.} Panel (a) corresponds to a semi-infinite geometry while panels (b) and (c) correspond to slab geometries, when both top and bottom surface Landau bands, represented by blue and green lines, are present. (a) Surface magnetization gaps the boundary spectrum of the semi-infinite TRS topological phase at $t=\pi$, as indicated by the bold vertical bar. As a result, the Weyl singularity is shifted.  (b) When the top and bottom surface magnetizations are parallel, the top and bottom Weyl singularities drift in opposite directions and the slab's Hall conductance at $t=\pi$ is 1. This corresponds to the case of an anomalous quantum Hall insulator. (c) When the top and bottom surface magnetizations are anti-parallel, the top and bottom Weyl singularities drift in the same direction, resulting in a net Hall conductance $0$ at $t=\pi$. This corresponds to the case of an axion insulator.}
\label{Fig:MagneticFilms}
\end{figure}

The discussion of the previous section can be extended to the experimental setting where the top and bottom surfaces of a TRS topological slab are gapped by ferromagnetic films. The magnetization of the films on the opposite surfaces of the slab can be parallel or anti-parallel. These two cases correspond to the quantum anomalous Hall insulator \cite{Chang2013, CYTT2014} and to the axion insulator \cite{XJS2018, MKT2017}, respectively, both already observed in laboratories. Our goal is to explain these effects using Corollary~\ref{Cor:Landau} and without invoking any half-integer Hall quantization.

\vspace{0.2cm}

We start the analysis with the anomalous Hall insulators. While the analysis is about a TRS insulator, it is actually useful to imagine an adiabatic process which connects this TRS-stabilized topological insulator with a trivial one. The ferromagnetic films act as small perturbations, more precisely, as local magnetic fields that open gaps in both top/bottom surface spectra of the TRS phase at $t=\pi$. As we learned in the previous sections, the topological gap-closings along an adiabatic cycle cannot be removed. What must have happened is that the topological gap-closing at $t=\pi$ is shifted by some $\delta$ along the $t$ axis, as illustrated in Fig.~\ref{Fig:MagneticFilms}(a). When the local magnetic fields are turned on, the familiar Landau bands form. Furthermore, if the magnetization of the top/bottom surfaces are parallel, the top/bottom Weyl singularities are shifted in opposite directions, as illustrated in Fig.~\ref{Fig:MagneticFilms}(b), and a spectral gap for the whole slab emerges at $t=\pi$. A counting similar to that in Fig.~\ref{Fig:HalfIQHE} gives $\tilde \sigma_H =1$, hence an anomalous quantum Hall insulator is observed at $t = \pi$.

\vspace{0.2cm}

The same argument can be repeated for axion insulators. Suppose the top and bottom magnetizations of the thin films are anti-parallel. In this case, both the top and bottom surface Weyl singularities shift in the same direction along the $t$ axis. As shown in Fig.~\ref{Fig:MagneticFilms}(b), a gap opens again in the spectrum of the slab at $t=\pi$, but the total Hall conductance is zero this time.

\section{Transfer of surface Hall States}
\label{Sec:TransferHall}

\begin{figure}
\center
\includegraphics[width=0.6\linewidth]{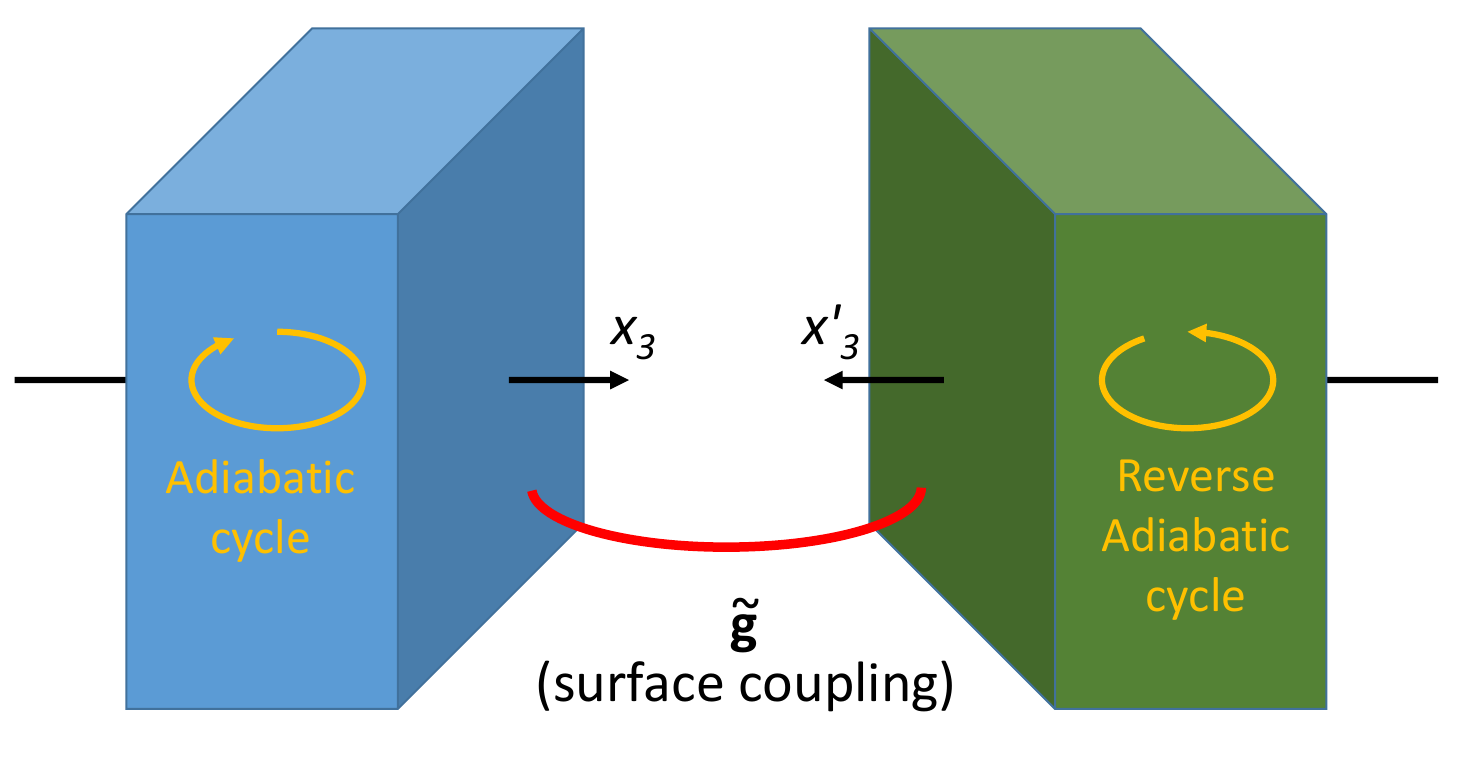}
\caption{\small {\bf Physical process described by the index map.} Two halved systems are driven in adiabatic cycles such that one cycle is time-reversed to the other. Their surfaces are put in proximity to each other during the cycle $t \in [0, 2\pi]$ and, as a result, a surface-surface coupling $\tilde{\bm g}$ is induced.}
\label{Fig:ChargeTransfer}
\end{figure}

Finally, we explore the surface physics by going all the way to $K_0(\Aa_3)$ in the diagram \eqref{Eq:KDiagram}. The main technical achievement of this section is the computation of the connecting map, which led us to the physical experiment illustrated in Fig.~\ref{Fig:ChargeTransfer}. It involves two copies of halved insulators driven in opposite adiabatic cycles. The surfaces of these two copies are coupled by a surface-to-surface potential
\begin{equation}
\begin{pmatrix} 0 & \tilde{g}(t) \\ \tilde{g}(t)^\ast & 0\end{pmatrix}, \quad \tilde{g}(t) \in \KM \otimes \widetilde \Aa_3, \quad t \in \SM^1,
\end{equation}
where $\KM$ is the algebra of compact operators over $\ell^2(\ZM^2 \times \NM)$. As a result, we have an adiabatic cyclic evolution of the combined systems driven by the Hamiltonian
\begin{equation}
\hat f(t) = \begin{pmatrix} \hat h(t) & \tilde{g}(t) \\ \tilde{g}(t)^\ast & \hat h(-t) \end{pmatrix}, \quad t \in \SM^1.
\end{equation}
The role of the surface-to-surface potential is to ensure that the spectra of $\hat f(t)$ are all gapped at $\bm G$. This scenario doesn't need fine tuning because it generically occurs when the surfaces are placed in close proximity to each other. Note that, because the two cycles are driven in the opposite directions on $\SM^1$, the total bulk second Chern number of the two copies is zero, hence there is no topological obstruction and the boundary spectrum can be gapped for all $t \in \SM^1$. The computation of the index map will show that quantum Hall states are transferred from one surface to the other when the systems are taken in a nontrivial adiabatic cycle with $\Delta \alpha_{\rm ME} \neq 0$. 

\subsection{Computation of the connecting map ${\rm Ind} \circ \theta^{-1}$}
\label{Sec:IndexMap}

The key to the computation is the notion of monodromy of a projection. Since it will be used in several contexts, we provide a generic definition below.

\begin{definition}\label{Def:Monodromy} For a $C^\ast$-algebra $\Aa$ and a projection $\bm p=\{p(t)\}_{t \in \SM^1} \in C(\SM^1,\Aa)$, we call the unique solution $v_t \in \Aa$ of the equation
\begin{equation}
i \partial_t v_t = i [ \partial_t p(t), p(t) ] v_t, \quad v_0 = p(0), \quad t \in \RM,
\end{equation}
the monodromy of $\bm p$. It is a partial isommetry satisfying the identities
\begin{equation}
v_t=v_t \, p(0) = p(t) \, v_t, \quad v_{2\pi}v_{2 \pi}^\ast = v_{2\pi}^\ast v_{2\pi}= p(0).
\end{equation} 
In particular, $v_{2\pi}+1- p(0)$ is a unitary element from $\Aa$.
\end{definition}

\begin{proposition}{\rm (\cite{PS}[Proposition.~4.3.7])} The inverse map $\theta^{-1}: K_0 (\sa_3) \to K_1 (\alg_3)$ is given by
\begin{equation}
\theta^{-1} \big ( [\bm p]_0  - [\bar{\bm p}]_0 \big ) = [v_{2\pi}+1-p(0)]_1,
\end{equation}
where $v_t$ is the monodromy of $\bm p \in \KM \otimes {\bm \Aa}_3$.
\end{proposition}

\begin{proposition}\label{Pro:IndMap}
Let $(\bm h,\bm G)$ be a gapped Hamiltonian from $\KM \otimes \bm \Aa_3$, $\hat{\bm h} \in \KM \otimes \hbA_3$ be any half-space Hamiltonian such that ${\rm ev}(\hat{\bm h}) = \bm h$, and suppose $\hat{ h}(0)$ is gapped at $\bm G$. Let $\bm h' \in \KM \otimes \bm \Aa_3$ be the element $\bm h'(t) =\bm h(-t)$ and $\hat{\bm h}'$ any element from $\KM \otimes \hbA_3$ such that ${\rm ev}(\hat{\bm h}')=\bm h'$. If $\tilde{\bm g}$ is any boundary element from $\KM \otimes \tbA_3$ such that
\begin{equation}
\hat{\bm f} = \begin{pmatrix} \hat{\bm h} & \tilde{\bm g} \\ \tilde{\bm g}^\ast & \hat{\bm h}' \end{pmatrix} \in \KM \otimes \hbA_3
\end{equation}
is gapped at $\bm G$ and $\tilde{g}(0)=0$, then
\begin{equation}\label{Eq:BBMap}
\big ({\rm Ind}\circ \theta^{-1}\big )\big ([\bm p_{\bm G}]_0 - [\bar{\bm p}_{\bm G}]_0 \big) = [\hat{p}_{2\pi}]_0 - [\hat{p}_0]_0,
\end{equation}
where
\begin{equation}\label{Eq:IndProj}
\hat{p}_{2\pi}=\hat w_{2\pi} \begin{pmatrix} \hat{p}_G(0) & 0 \\ 0 & 0 \end{pmatrix} \hat w_{2\pi}^\ast, \quad \hat{p}_{0}=\begin{pmatrix} \hat{p}_G(0) & 0 \\ 0 & 0 \end{pmatrix},
\quad \hat{\bm p}_{\bm G} = \chi_{(-\infty,\bm G]}(\hat{\bm h}),
\end{equation}
with $\hat w_t$ being the monodromy of the projector $\hat{\bm q}_{\bm G} = \chi_{(-\infty,\bm G]}(\hat{\bm f})$.
\end{proposition}

\proof According to the standard definition of the index map ${\rm Ind}: K_1(\Aa_3) \rightarrow K_0(\widetilde \Aa_3)$, we have \cite{RLL}[Ch.~9.1]
\begin{equation}\label{Eq:DefIndexMap}
{\rm Ind}([u]_1) = \left [\hat v \begin{pmatrix} 1 & 0 \\ 0 & 0 \end{pmatrix} \hat v^\ast \right]_0 - \left [ \begin{pmatrix} 1 & 0 \\ 0 & 0 \end{pmatrix} \right ]_0,
\end{equation}
where $u$ belongs to the group $\UM (\KM \otimes \Aa_3)$ of unitary elements from $\KM \otimes \Aa_3$, and $\hat v$ is a lift of the unitary operator ${\small \begin{pmatrix} u & 0 \\ 0 & u^\ast \end{pmatrix}}$ to an element belonging to $\KM \otimes \widehat \Aa_3$. In other words
\begin{equation}
{\rm ev}(\hat v) = \begin{pmatrix} u & 0 \\ 0 & u^\ast \end{pmatrix}.
\end{equation}
It is important to keep in mind that any lift will do, as the defining relation \eqref{Eq:DefIndexMap} is entirely independent of the choice of this lift. Furthermore, even though the elements appearing in \eqref{Eq:DefIndexMap} belong to the half-space algebra $\ha_3$, the $K$-element written in \eqref{Eq:DefIndexMap} belongs to $K_0(\widetilde \Aa_3)$.

\vspace{0.2cm}

From the previous proposition, the input for the index map in \eqref{Eq:BBMap} is $v_{2\pi}+1-p_G(0)\in \UM(\KM\otimes\Aa_3)$, where $v_t$ is the monodromy of $\bm p_{\bm G}$. While the lift is known to always exist \cite{RLL, WO}, there is no generic algorithm to construct it. As such, our main task is to find such lift that can be expressed entirely in terms of the given data. A key observation is that the monodromy $v'_t$ of $\bm p'_{\bm G} = \chi_{(-\infty,\bm G]}(\bm h')$ satisfies $v'_t = v_t^\ast$. Then our task reduces to finding a lift of
\begin{equation}\label{Eq:MonodromyMatrix}
\begin{pmatrix} v_{2\pi}+1 - p_G(0) & 0 \\ 0 & v'_{2\pi}+1 - p_G(0) \end{pmatrix} ,
\end{equation}
and it is here where $\hat{\bm f}$ enters the picture. Indeed, observe that
\begin{equation}
{\rm ev} (\hat{\bm f} )= \begin{pmatrix} \bm h & 0 \\ 0 & \bm h' \end{pmatrix},
\end{equation}
and since the map ${\rm ev}$ is an algebra endomorphism, it commutes with functional calculus. As such, we automatically have
\begin{equation}
{\rm ev}(\hat w_t) =  \begin{pmatrix} v_t & 0 \\ 0 & v'_t \end{pmatrix},
\end{equation}
with ${\hat w_t}$ as defined in the statement. Then the lift of ${\small \begin{pmatrix} v_{2\pi} & 0 \\ 0 & v_{2\pi}^\ast \end{pmatrix}}$ is given by
\begin{equation}
{\rm Lift}\begin{pmatrix} v_{2\pi} & 0 \\ 0 & v^\ast_{2\pi} \end{pmatrix} = \hat w_{2 \pi}.
\end{equation}
Since
\begin{equation}
{\rm Lift}\begin{pmatrix} 1- p_G(0) & 0 \\ 0 & 1 - p_G(0) \end{pmatrix} = 1 - \hat{q}_G(0),
\end{equation}
we have just found the lift of \eqref{Eq:MonodromyMatrix}.
Using the definition of the index map \eqref{Eq:DefIndexMap}, we obtain
\begin{align}
& \qquad \qquad \quad \big ({\rm Ind}\circ \theta^{-1}\big ) \big ([\bm p_{\bm G}]_0 - [\bar{\bm p}_{\bm G}]_0 \big) = \\ \nonumber 
& \left [\big (\hat w_{2 \pi}+1-\hat{q}_G(0)\big ) {\small \begin{pmatrix} 1 & 0 \\ 0 & 0 \end{pmatrix}} \big (\hat w_{2 \pi}+1-\hat{q}_G(0)\big )^\ast \right]_0 - \left [ {\small \begin{pmatrix} 1 & 0 \\ 0 & 0 \end{pmatrix}} \right ]_0.
\end{align}
Since $\tilde{g}(0)=0$, the following simplification occurs:
\begin{equation}
1 - \hat{q}_G(0) = \begin{pmatrix} 1-\hat{p}_G(0) & 0 \\ 0 & 1 - \hat{p}_G(0) \end{pmatrix}
\end{equation}
Hence, $1-\hat{q}_G(0)$ commutes with ${\small \begin{pmatrix} 1 & 0 \\ 0 & 0 \end{pmatrix}}$, and we have the relation $\hat w_{2 \pi} \big ( 1-\hat{q}_G(0)\big )=0$. The statement follows. \qed

\vspace{0.2cm}

Let us follow up with a simple application of the above result. In the previous section, we concluded that there cannot be a global gap $\hat G(t)$ in the spectrum of $\hat h(t)$, and the argument was based on the spectral flow. We now have a second route to arrive at the same conclusion.

\begin{corollary}\label{Cor:IndMap} If
\begin{equation}
\Delta \alpha_{\rm ME} = {\rm Ch}_{\{0,1,2,3\}}(\bm p_{\bm G}) \neq 0,
\end{equation} 
then necessarily ${\rm Spec}(\hat{\bm h})\cap \bm G = \bm G$.
\end{corollary}

\proof If ${\rm Spec}(\hat{\bm h})\cap \bm G$ is not the full $\bm G$, then $\hat{\bm h}$ is gapped at $\bm G$ and we do not need $\tilde{\bm g}$ in the expression of $\hat{\bm f}$. In this case, the lift involved in the index map is trivial and as an immediate consequence $({\rm Ind}\circ \theta^{-1})\big ([\bm p_{\bm G}]_0 - [\bar{\bm p}_{\bm G}]_0 \big)  = [0]_0$. On the other hand, the action of the connecting maps on the generators of the $K$-groups are known explicitly \cite{PS}[Sec.~4.2.3]. In particular, the top generator $[\bm e_{\{0,1,2,3\}}]_0$ of the $K_0$-group of ${\bm \Aa}_3 \simeq \Aa_4$ is mapped by ${\rm Ind}\circ \theta^{-1}$ onto $[\tilde{e}_{\{1,2\}}]_0 \in K_0(\widetilde{\Aa}_3)$. By assumption $[\bm p_{\bm G}]_0 - [\bar{\bm p}_{\bm G}]_0$ contains the top generator, hence, $\big ({\rm Ind}\circ \theta^{-1}\big )\big ([\bm p_{\bm G}]_0 - [\bar{\bm p}_{\bm G}]_0 \big) \neq [0]_0$. Therefore, ${\rm Spec}(\hat{\bm h})\cap \bm G$ needs to cover $\bm G$ entirely.\qed 

\begin{figure}
\center
\includegraphics[width=0.6\linewidth]{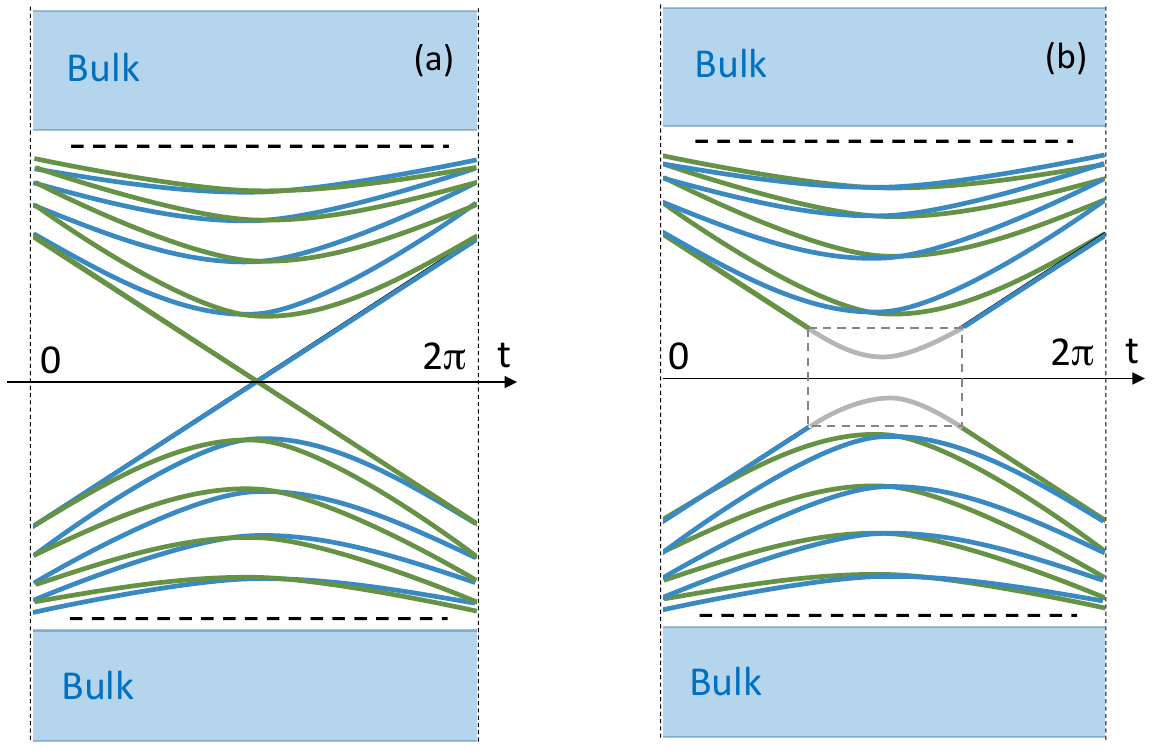}
\caption{\small {\bf The flow of the Landau levels of $\hat{\bm f}$ with $\tilde{\bm g}$ (a) turned off and (b) turned on.} (a). When the surface-surface coupling $\tilde{\bm g}$ is turned off, two halved systems act as independent systems. Their chiral Landau flows of $\bm h$ and $\bm h'$ are represented by the blue and green lines. The Landau flow carries Hall states from valence band to conduction band for the blue system, and vice versa for the green system. (b). Due to the coupling $\tilde{\bm g}$, a hybridization (in the grey region) is formed between the blue and green Landau bands and adiabatic transfer of Hall states is now possible between the blue and green systems.}
\label{Fig:LandaFlow3}
\end{figure}

\subsection{Physical interpretation} 

We now return to the physical systems described at the beginning of the section and illustrated in Fig.~\ref{Fig:ChargeTransfer}. Suppose we prepare the systems in an initial state by populating the entire spectrum below the gap $\hat G(0)$ of $\hat h(0)$ describing the blue system. This is possible because the blue and the green systems are fully decoupled at $t=0$. The adiabatic theorem assures us that the state of the blue system evolves, after a full adiabatic cycle, into the state defined by the projection $\hat p_{2\pi}$ defined in \eqref{Eq:IndProj}. According to Proposition~\ref{Pro:IndMap} and Corollary~\ref{Cor:IndMap}, if $\Delta \alpha_{\rm ME} \neq 0$, then the two quantum states belong to different $K_0$-classes of $\KM \otimes \widehat \Aa_3$. Furthermore, the difference between the two classes is an element of $K_0(\widetilde \Aa_3) \simeq K_0(\Aa_2)$. Hence
\begin{equation}\label{Eq:C1}
[\hat p_{2\pi}]_0 - [\hat p_0]_0 = n[1]_0 + m [\tilde e_{\{1,2\}}]_0,
\end{equation}
and for the element on the right we can define a boundary Chern number as explained in section~\ref{Sec:HSAlgebra}. The Chern number is a group morphism from $K_0$ to the real axis and on the generators of $K_0(\widetilde \Aa_3)$ it gives $\widetilde {\rm Ch}_{\{1,2\}}\big ( [\tilde e_{\{1,2\}}]_0 \big ) =1$ and $\widetilde {\rm Ch}_{\{1,2\}}\big ( [1]_0 \big ) = 0$. When it is evaluated on the right hand side of \eqref{Eq:C1}, we get $m$. Combining \cite{PS}[Theorem.~5.4.1] and \cite{PS}[Theorem.~5.5.1], we find
\begin{equation}
m = \widetilde{\rm Ch}_{\{1,2\}}\Big ( \big ({\rm Ind}\circ \theta^{-1}\big ) \big ([\bm p_{\bm G}]_0 - [\bar{\bm p}_{\bm G}]_0 \big ) \Big ) = -{\rm Ch}_{\{0,1,2,3\}}(\bm p_G) = -\Delta \alpha_{\rm ME}.
\end{equation}
The conclusion is that surface states carrying a boundary Chern number equal to $-\Delta \alpha_{\rm ME}$ have been pumped from continuum spectrum. Since all the blue states are already fully populated and the system remained gapped throughout the adiabatic cycle, the surface Chern states of the blue system must have been transferred to the surface of the green system. 

\vspace{0.2cm}

To see this phenomenon more directly, let us introduce a magnetic field perpendicular to the surfaces of the two copies, such that the surface spectrum is resolved in Landau levels. The flow of the Landau levels in the absence of the coupling is shown in Fig.~\ref{Fig:LandaFlow3}(a). It was obtained by super imposing the flow from Fig.~\ref{Fig:SpecFlow1} over its mirror image, as it should for the two identical systems in Fig.~\ref{Fig:ChargeTransfer}. When the coupling is introduced, the blue and green Landau levels can hybridize and avoided crossings can emerge. We can design a specific coupling which produces only one avoided crossing as illustrated in Fig.~\ref{Fig:LandaFlow3}(b). For this, let
\begin{equation}
p_{\Delta E}(t) = \chi_{[E_-,E_+]} \big (\hat h(t) \big )
\end{equation}
be the spectral projector of the blue system onto the energy interval $[E_-,E_+]$ shown in Fig.~\ref{Fig:LandaFlow3}(b). Then we take the coupling as
\begin{equation}\label{Eq:Coupling1}
\tilde s(t)=\gamma(t) \begin{pmatrix} 0 & p_{\Delta E}(t) \, p_{\Delta E}(-t) \\ p_{\Delta E}(-t) \, p_{\Delta E}(t) & 0 \end{pmatrix},
\end{equation}
with $\gamma(t)$ a smooth function with support in the interval indicated in Fig.~\ref{Fig:LandaFlow3}(b). This choice makes $\tilde s(t)$ smooth in $t$. If
\begin{equation}
\hat{\bm f}_0 = \begin{pmatrix} \hat{\bm h} & 0 \\ 0 & \hat{\bm h}' \end{pmatrix}
\end{equation}
is the adiabatic family of Hamiltonians for the uncoupled systems, the spectral projector of $\hat{\bm f}_0$ onto the interval $[E_-,E_+]$ is just
\begin{equation} 
\chi_{[E_-,E_+]} \big (\hat f_0(t) \big ) = \begin{pmatrix} p_{\Delta E}(t) & 0 \\ 0 & p_{\Delta E}(-t) \end{pmatrix},
\end{equation} 
which commutes with $\tilde s(t)$,
\begin{equation}
\chi_{[E_-,E_+]} \big (\hat f_0(t) \big ) \tilde s(t) = \tilde s(t) \chi_{[E_-,E_+]} \big (\hat f_0(t) \big ).
\end{equation}
This assures us that only the states inside the spectral interval $[E_-,E_+]$ will be affected by $\tilde s(t)$. By construction, there are only two Landau levels inside this interval, hence the hybridization and the avoided crossing must be as shown in Fig.~\ref{Fig:LandaFlow3}(b).

\vspace{0.2cm}

Once the spectral flow in Fig.~\ref{Fig:LandaFlow3}(b) is established, we can easily follow the adiabatic evolution of the Landau levels. Recall that we start at $t=0$ with all the blue Landau levels populated and the green ones empty. We can see that the first blue Landau level evolves into the first green Landau level, counting from the gap, while the rest of the blue Landau levels fully re-populate the blue states at the end of the adiabatic cycle. The only net difference is the populated top green Landau level at the end of the adiabatic cycle.

\vspace{0.2cm}

In general, the number of Landau levels transferred to the green system will depend on the surface-to-surface coupling and the coupling can vary drastically from one sample to another. However, Proposition~\ref{Pro:IndMap} assures us that the number of populated green levels minus the number of un-populated blue levels is independent of the coupling and is equal to $-\Delta \alpha_{\rm ME}$. Using the flows of Landau levels, the statement can be shown to remain the same even if the quantum states of the green system are partially populated instead of being empty. 

\subsection{Experimental predictions II}

We list our experimental predictions based on the index map calculations:

\begin{itemize}

\item Suppose in the lab one has two copies of the same topological insulator, both halved and coupled as above. If a bias potential is maintained between the samples such that a population imbalance is produced between the two copies, then a change of the total surface Hall conductance equal to $\Delta \alpha_{\rm ME}$ will be observed after each adiabatic cycle.

\item The surface of an insulator can be prepared in a topological Chern state, which is generated by a non-zero second Chern number via the proximity effect described above. Having the quantum states of the second system fully depopulated is impossible for condensed matter systems, but this proposal can be tested with photonic and mechanical topological insulators \cite{BarlasPRB2018}.

\end{itemize}

\ack 
E. P. acknowledges financial support from US National Science Foundation through grant DMR-1823800.

\end{document}